\documentclass[twocolumn,superscriptaddress,amsmath,amssymb,aps,longbibliography,nofootinbib,prx]{revtex4-2}
\usepackage{amssymb}
\usepackage{graphicx}
\usepackage{booktabs}
\usepackage{amsmath}
\usepackage{multirow}
\usepackage{blkarray}
\usepackage{nicematrix}
\usepackage{tikz}
\usepackage{mathrsfs}
\usepackage{hyperref}
\hypersetup{colorlinks,breaklinks,
            urlcolor=[rgb]{0.25,0.41,0.88},
            linkcolor=[rgb]{0.25,0.41,0.88},
            citecolor=[rgb]{0.25,0.41,0.88}}
\usepackage{braket}
\usepackage{babel}
\usepackage{blindtext}
\usepackage{soul}
\usepackage[compat=1.1.0]{tikz-feynman}
\usepackage[normalem]{ulem}
\usepackage{physics}

\usepackage{mathtools}
\usepackage{bbm}
\usepackage{mathtools}

\definecolor{mypurp}{rgb}{0.35, 0, 0.7}

\usepackage{amsthm}
\usepackage{txfonts}

\theoremstyle{definition}

\newcommand{\SPT}{\text{SPT}}
\newcommand{\gSPT}{\text{gSPT}}
\newcommand{\gtri}{\text{gTri}}

\newcommand{\free}{\text{free}}
\newcommand{\fix}{\text{fix}}
\newcommand{\even}{\text{e}}
\newcommand{\odd}{\text{o}}

\newcommand{\CZ}{\text{CZ}}

\newcommand{\ent}{\text{ent}}

\newcommand{\Ising}{\text{Ising}}

\newcommand{\clk}{\text{clk}}
\newcommand{\LL}{\text{LL}}

\newcolumntype{C}[1]{>{\centering\arraybackslash}p{#1}}
\begin{document}

\def\papertitle{A Framework for Predicting Entanglement Spectra of \\Gapless Symmetry-Protected Topological States in One Dimension}

\newcommand{\TUM}{\affiliation{Technical University of Munich, TUM School of Natural Sciences, Physics Department, 85748 Garching, Germany}}
\newcommand{\MCQST}{\affiliation{Munich Center for Quantum Science and Technology (MCQST), Schellingstr. 4, 80799 M{\"u}nchen, Germany}}

\newcommand{\XJTU}{\affiliation{Shaanxi Province Key Laboratory of Advanced Materials and Mesoscopic Physics, Key Laboratory for Nonequilibrium Synthesis and
Modulation of Condensed Matter of Ministry of Education, School of Physics, Xi'an Jiaotong University, Xi'an 710049, China}}

\author{Wen-Tao Xu} 
\XJTU\TUM \MCQST
\author{Frank Pollmann}  \TUM \MCQST
\author{Michael Knap}  \TUM \MCQST

\title{\papertitle}

\begin{abstract}
The concept of gapped symmetry-protected topological (SPT) states has been generalized to gapless SPT (gSPT) states. 
Similar to gapped SPT states, gSPT states in one dimension exhibit universal degeneracies in their entanglement spectra. 
The entanglement spectra of gSPT states are further described by boundary conformal field theories, whose systematic prediction is a key open question. 
To address this problem, we focus on the class of gSPT states that are obtained by applying unitary SPT entanglers to trivial, critical states in one dimension. 
We find that the reduced density matrix of a non-trivial gSPT state can be obtained, either exactly or approximately, by applying a quantum channel to the reduced density matrix of the trivial gSPT state. 
This quantum channel acts only near the entanglement cut and modifies its corresponding conformal boundary condition, allowing us in turn to predict the boundary conformal field theory describing the entanglement spectra. 
We apply this framework to gSPT states protected by various symmetries and having different central charges, and further analyze the stability of boundary conditions of the entanglement cut. 
Our work thereby provides a framework for systematically analyzing and understanding the entanglement spectra of gSPT states.         
\end{abstract}
\maketitle
\tableofcontents
\section{Introduction}

The concept of symmetry-protected topological (SPT) phases has significantly evolved our understanding of gapped, disordered quantum phases in the presence of symmetries~\cite{haldane_1981,Gu_Wen_2009,Pollmann_ES_2010,SPT_science_2012,Pollmann_2012,SPT_cohomology_2013,SPT_Calssify_1D_2011,SPT_complete_classify_1D_2011,Classify_schuch_2011}. These phases are gapped and cannot be adiabatically connected to trivial phases without either breaking the protecting symmetry or closing the energy gap~\cite{LU_LRE_2010}. Gapped SPT phases can be characterized by nonlocal string order parameters~\cite{den_Nijs_string_1989,KT_1992,Pollmann_SOP_2012,SOP_jutho_2012}, protected edge modes~\cite{CZX_2011,Levin_GU_2012,Olexei_Motrunich_2014}, and characteristic degeneracies in the entanglement spectrum (ES)~\cite{Li_Haldane_2008,Pollmann_ES_2010}. From a practical perspective, some SPT states can serve as resources for measurement-based quantum computation due to their distinctive entanglement structure~\cite{MBQC_2012,MBQC_njp_2012}. As a result, SPT phases have become one of the most active research topics in condensed matter physics and quantum information. 

\begin{figure}
     \centering
    \includegraphics[width=\linewidth]{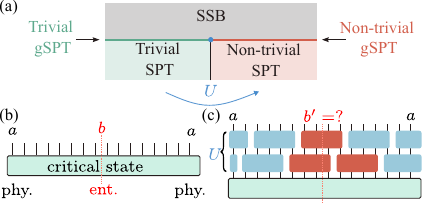}
    \caption{
    \textbf{Concept of  gapless Symmetry Protected Topological (gSPT) states and their construction.}
    (a) We focus on non-intrinsic gSPT states realized at critical points separating gapped SPT phases from symmetry-breaking phases that spontaneously break the protecting symmetries.  non-trivial gSPT (SPT) states are obtained by applying SPT entanglers $U$ to trivial gSPT (SPT) states. (b) For a 1D critical state, the ES is determined by the physical boundary condition $a$ and the boundary condition $b$ of the entanglement cut (entanglement boundary condition). (c) After applying an SPT entangler to the critical state, the entanglement boundary condition is generally modified to an a priori unknown boundary condition. In the illustration, the red gates modify the ES while blue ones do not. 
}
    \label{fig:problem}
\end{figure}

In recent years, it has been shown that gapped SPT phases can be generalized to gapless SPT (gSPT) phases~\cite{Scaffidi_gSPT_2017,Ruben_2018,Gapless_SPT_Ruben_2021,Intrinsic_gSPT_ruben_2021}. Distinct gSPT phases cannot be connected to one another along symmetric paths governed by the same underlying conformal field theory (CFT). Depending on whether a gapped counterpart exists, gSPT states can be divided into two classes. Non-intrinsic gSPT states possess a corresponding gapped SPT phase, whereas intrinsic gSPT states do not~\cite{Gapless_SPT_Ruben_2021,KT_4_gSPT_2025,XJ_Xu_reviwe_2026}.
In this work, we focus on non-intrinsic gSPT states, which typically arise at critical boundaries between gapped SPT phases and symmetry-breaking phases that break the protecting symmetries, as illustrated in Fig.~\ref{fig:problem}a. The characterization of non-intrinsic gSPT phases using edge modes and string order parameters has been explored~\cite{Gapless_SPT_Ruben_2021}. Although the entanglement spectra (ES) of one-dimensional gSPT states have been studied via bulk–boundary correspondence~\cite{Li_Haldane_2008,Yu_2025}, it remains open how to systematically predict and analyze the ES of such gSPT states.

Distinct from the ES of one-dimensional gapped SPT states, which exhibits characteristic degeneracies protected by the underlying SPT order~\cite{Pollmann_ES_2010}, the ES of one-dimensional gSPT states displays richer universal structures. In addition to SPT order induced degeneracies, it is described by boundary conformal field theories (BCFT)~\cite{CARDY_1984,CARDY_1986,CARDY_1989,CARDY_2006}, as is typical for the ground state ES of critical spin chains~\cite{lauchli_ES_BCFT_2013}. The ES of a one-dimensional critical state is determined by two boundary conditions, as illustrated in Fig.~\ref{fig:problem}b: the physical boundary condition of the spin chain (for simplicity, we assume identical physical boundary conditions at both ends) and the entanglement boundary condition induced by the entanglement cut. Numerical studies have shown that for conventional critical spin chains—i.e., systems without Hilbert-space constraints or non-trivial SPT order—the entanglement boundary condition is typically the free boundary~\cite{Roy_Saleur_Pollmann_2020,Roy_ent_boundary_2025}. Interestingly, recent numerical results indicate that the entanglement boundary conditions for trivial and non-trivial gSPT states can differ~\cite{Huang_2024}. This raises a central question: how can one systematically understand and predict the entanglement boundary conditions as well as the ES of one-dimensional non-trivial gSPT phases?

To address this problem, we focus on non-intrinsic gSPT states. Because these states admit gapped counterparts that can be constructed by applying unitary SPT entanglers to trivial gapped states~\cite{SPT_cohomology_2013,Santos_2015,Pivot_2023,Carolyn_Zhang_2023}, their gapless analogs in one dimension can be similarly obtained by applying the same entanglers to critical states described by CFTs, as illustrated in Fig.~\ref{fig:problem}a. Based on this construction, we show how the reduced density matrix of a non-trivial gSPT state is obtained either exactly or approximately by applying a quantum channel to the reduced density matrix of the corresponding trivial gapless state. Furthermore, we find from many examples that the Kraus operators of this quantum channel can be expressed in terms of projectors, which allows us to understand how the entanglement boundary condition of a critical state is changed by the action of an SPT entangler. By invoking the bulk–boundary correspondence~\cite{Li_Haldane_2008}, we determine the effective entanglement Hamiltonians (EH) and predict the corresponding ES. Using this framework, we analyze the ES of several representative examples, including gSPT states protected by on-site unitary symmetries, antiunitary time-reversal symmetry, and non-invertible symmetries. Additionally, we discuss the stability of the associated conformal boundary conditions characterizing the entanglement boundaries.

This paper is organized as follows. In Sec.~\ref{sec:c_half_z2_z2}, we present a warm-up example by studying 1D $\mathbb{Z}_2 \times \mathbb{Z}_2$ gSPT states with central charge $c = 1/2$. We show that their reduced density matrices are exactly related by a quantum channel and analyze their ES. In Sec.~\ref{sec:framework}, we introduce a general framework to analyze how SPT entanglers modify the entanglement boundary and ES. In Sec.~\ref{Sec:Z_2T_and_c_half}, we demonstrate that our framework remains valid even when the reduced density matrices are only approximately related by a quantum channel, using $\mathbb{Z}_2 \times \mathbb{Z}_2^T$ gSPT states with central charge $c = 1/2$ as an example. In Sec.~\ref{Sec:c=1_Z2_Z2}, we apply the framework to $\mathbb{Z}_2 \times \mathbb{Z}_2$ gSPT states with central charge $c = 1$ and a continuously varying Luttinger parameter. In Sec.~\ref{sec:non_inv_SPT}, we apply our framework to non-invertible SPT and gSPT states realized by the group-based cluster states. Sec.~\ref{sec:conclusion_discusion} contains conclusions and discussions. Additional details of the framework and further examples are provided in the Appendices.

\section{Illustrative example: ES of $\mathbb{Z}_2\times\mathbb{Z}_2$ gSPT states with central charge $c=\frac{1}{2}$}\label{sec:c_half_z2_z2}

One of the simplest 1D gSPT states is the $\mathbb{Z}_2\times\mathbb{Z}_2$ gSPT state with the central charge $c=1/2$~\cite{Scaffidi_gSPT_2017}. The Hamiltonian for the trivial gSPT model is
\begin{equation}\label{eq:Z2_Z2_c_half_gtri}
    H_{\gtri}=-\sum_{n=1}^{N-2}Z_{2n}Z_{2n+2}-\sum_{n=1}^{2N-1} X_n,
\end{equation}
where $\{X,Y,Z\}$ denote the Pauli matrices. It has a $\mathbb{Z}_2\times \mathbb{Z}_2$ symmetry generated by $\prod_{n=1}^{N-2} X_{2n}$ and $\prod_{n=1}^{N} X_{2n-1}$, which act separately on the even (e) and odd (o) sublattices. On the even sublattice, the Hamiltonian reduces to the critical transverse-field Ising chain with free open boundary condition (OBC) at both ends (see detailed definitions of various OBC of the Ising chain in Appendix~\ref{app:OBC_IMPS}), and we denote its ground state as $\ket{\Psi_\Ising}_{\text{e}}$. On the odd sublattice, the Hamiltonian is: $-\sum_n X_{2n+1}$, and its ground state is $\ket{+}_{\odd} = \prod_{n=1}^N \ket{+}_{2n-1}$ with $\ket{+} = (\ket{0} + \ket{1})/\sqrt{2}$. So the ground state of $H_{\gtri}$ is
$\ket{\Psi_\gtri} = \ket{+}_{\odd} \otimes \ket{\Psi_\Ising}_{\even}$.

By applying the SPT entangler $U_{CZ}=\prod_{n=1}^{2N-2}CZ_{n,n+1}$ to $\ket{\Psi_{\gtri}}$, where $CZ_{n,m}=\frac{1+Z_n}{2}+\frac{1-Z_n}{2}Z_m$ is the controlled-Z gate, the non-trivial gSPT state is obtained: $\ket{\Psi_\gSPT}=U_{\CZ}\ket{\Psi_\gtri}$~\cite{Scaffidi_gSPT_2017}. Since $U_{CZ}$ commutes with the $\mathbb{Z}_2\times\mathbb{Z}_2$ symmetry generators in the bulk (which is best seen on a circuit level when pulling the generators through $U_{CZ}$), the non-trivial gSPT state also has the $\mathbb{Z}_2\times \mathbb{Z}_2$ symmetry. 
To investigate the general properties of the ES in the non-trivial gSPT phase, we generalize the SPT entangler $U_{CZ}$ to $U_{CZ}U_X(\theta)$, where $U_X(\theta)=\prod_{n=1}^{2N-1}\exp(i\theta X_n)$ also preserves the $\mathbb{Z}_2\times\mathbb{Z}_2$ symmetry. The non-trivial gSPT state is then generalized to 
\begin{align}\label{eq:psi_gSPT}
    \ket{\Psi_\gSPT(\theta)} &= U_{\CZ}U_X(\theta)\ket{\Psi_\gtri}.
\end{align}
And its Hamiltonian is
\begin{equation}
    H_{\text{gSPT}}(\theta) = U_{CZ} U_X(\theta) H_{\gtri} U_X^\dagger(\theta) U_{CZ}^\dagger.
\end{equation}
Because
\begin{equation}
    U_{CZ}U_{X}(\theta) = \prod_n \exp(i\theta Z_{n-1} X_n Z_{n+1}) U_{CZ}
\end{equation}
and the gate $\exp(i\theta Z_{n-1} X_n Z_{n+1})$ is $\mathbb{Z}_2\times\mathbb{Z}_2$ symmetric, the family of states $\ket{\Psi_\gSPT(\theta)}$ all belong to the same $\mathbb{Z}_2\times\mathbb{Z}_2$ gSPT phase.

Since $\ket{\Psi_\gSPT(\theta)}$ and $\ket{\Psi_\gtri}$ are related via the SPT entangler, one natural question is how their reduced density matrices are related. 
Consider an entanglement cut in the middle of the chain, i.e., between sites $N-1$ and $N$, so that subsystem $A$ consists of sites $1$ to $N-1$, and subsystem $B$ consists of sites $N$ to $2N-1$. 
We find that the reduced density matrices 
$
\rho_{\gtri} = \Tr_A \ket{\Psi_{\gtri}}\bra{\Psi_{\gtri}} \quad \text{and} \quad 
\rho_{\gSPT}(\theta) = \Tr_A \ket{\Psi_{\gSPT}(\theta)}\bra{\Psi_{\gSPT}(\theta)}
$
are exactly related via a quantum channel:\footnote{The general relation between reduced density matrices is $
     \rho_{\gSPT}(\theta)\cong\mathscr{N}_{CZ}[\mathscr{N}_X[\rho_{\gtri}]]+\frac{1}{2}\left(\mathscr{N}_X[\rho'_{\gtri}]-Z_N\mathscr{N}_X[\rho'_{\gtri}]Z_N\right)$
where $\rho'_{\gtri}=\Tr_A\left(Z_{N-1}\ket{\Psi_{\gtri}}\bra{\Psi_{\gtri}}\right)$. Since $\ket{\Psi_{\gtri}}=\ket{+}_{\odd}\otimes\ket{\Psi_{\Ising}}_\even$,  $\rho'_{\gtri}=0$ and we have Eq.~\eqref{eq:channel4CZX}.}
\begin{align}\label{eq:channel4CZX}
    \rho_{\gSPT}(\theta)&\cong \varrho_{\gSPT}(\theta)=\mathscr{N}_{CZ}\left[\mathscr{N}_{X}[\rho_{\gtri}]\right],\notag\\
        \mathscr{N}_{CZ}[\rho]&=\frac{1}{2}\left(\rho+Z_N\rho Z_N\right) \text{ and } \mathscr{N}_X[\rho]=e^{i\theta X_N}\rho e^{-i\theta X_N}
\end{align}
where ``$\cong$'' means that we ignore unitary or isometric gates that do not affect the ES, as illustrated in Fig.~\ref{fig:problem}c. Evidence that $\ket{\Psi_\gSPT(\theta)}$ is a non-trivial gSPT state can be obtained from the projective representation of the protecting symmetry acting on its reduced density matrix. One finds that $\varrho_{\gSPT}(\theta)$ carries a projective $\mathbb{Z}_2\times\mathbb{Z}_2$ symmetry generated by $Z_N X_{N+1} X_{N+3} \cdots X_{2N-1}$ and $X_N X_{N+2} X_{N+4} \cdots X_{2N-2}$, which anti-commute with each other, implying that $\ket{\Psi_{\gSPT}(\theta)}$ realizes a non-trivial gSPT state.

\begin{figure}
     \centering
    \includegraphics[width=0.7\linewidth]{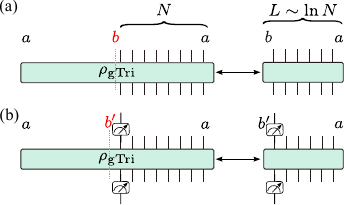}
    \caption{\textbf{The effective Entanglement Hamiltonian (EH) from the bulk--boundary correspondence.} 
(a) For a critical state with physical boundary condition $a$ and entanglement boundary condition $b$, the low-lying ES is analogous to the low-energy spectrum of an OBC chain whose right (left) boundary condition is $a$ ($b$). This OBC chain can be viewed as an effective EH, whose length scales logarithmically with the subsystem size $N$.
(b) The entanglement boundary condition is modified by the Kraus projectors of the quantum channel induced by the SPT entangler, Eq.~\eqref{eq:kraus_proj_Z2_Z2}, and the quantum channel is implemented using a projective measurement. As a result, the left boundary condition of the effective EH is changed accordingly.}
    \label{fig:bulk_boundary}
\end{figure}

We now turn to the ES of the $\mathbb{Z}_2\times\mathbb{Z}_2$ gSPT states. Since $\ket{\Psi_{\gtri}}=\ket{+}_{\odd}\otimes\ket{\Psi_{\Ising}}_{\even}$, its ES is identical to that of the critical Ising chain with free physical OBC. Because the entanglement boundary condition of the Ising chain is also free, a phenomenological low-energy effective EH for $\ket{\Psi_{\gtri}}$ can be inferred from the bulk--boundary correspondence~\cite{Li_Haldane_2008,Cho_ES_BCFT_PRB_2017,Huang_2024,Universal_ES_2024}:
\begin{equation}\label{eq:H_E_Ising}
    H_{E,\gtri}\sim-\sum_{n=1}^{L-1}Z_{n}Z_{n+1}-\sum_{n=1}^{L}X_{n},
\end{equation}
whose length scales logarithmically with the subsystem size $L\sim\ln(N)$~\cite{Calabrese_Cardy_2004,Calabrese_cardy_2009}.  The effective EH may live on a lattice different from the original one, so we relabel the lattice sites accordingly. Here the left (right) boundary of the EH corresponds to the entanglement (physical) boundary, as illustrated in Fig.~\ref{fig:bulk_boundary}a.

\begin{figure}
    \centering
        \includegraphics[width=\linewidth]{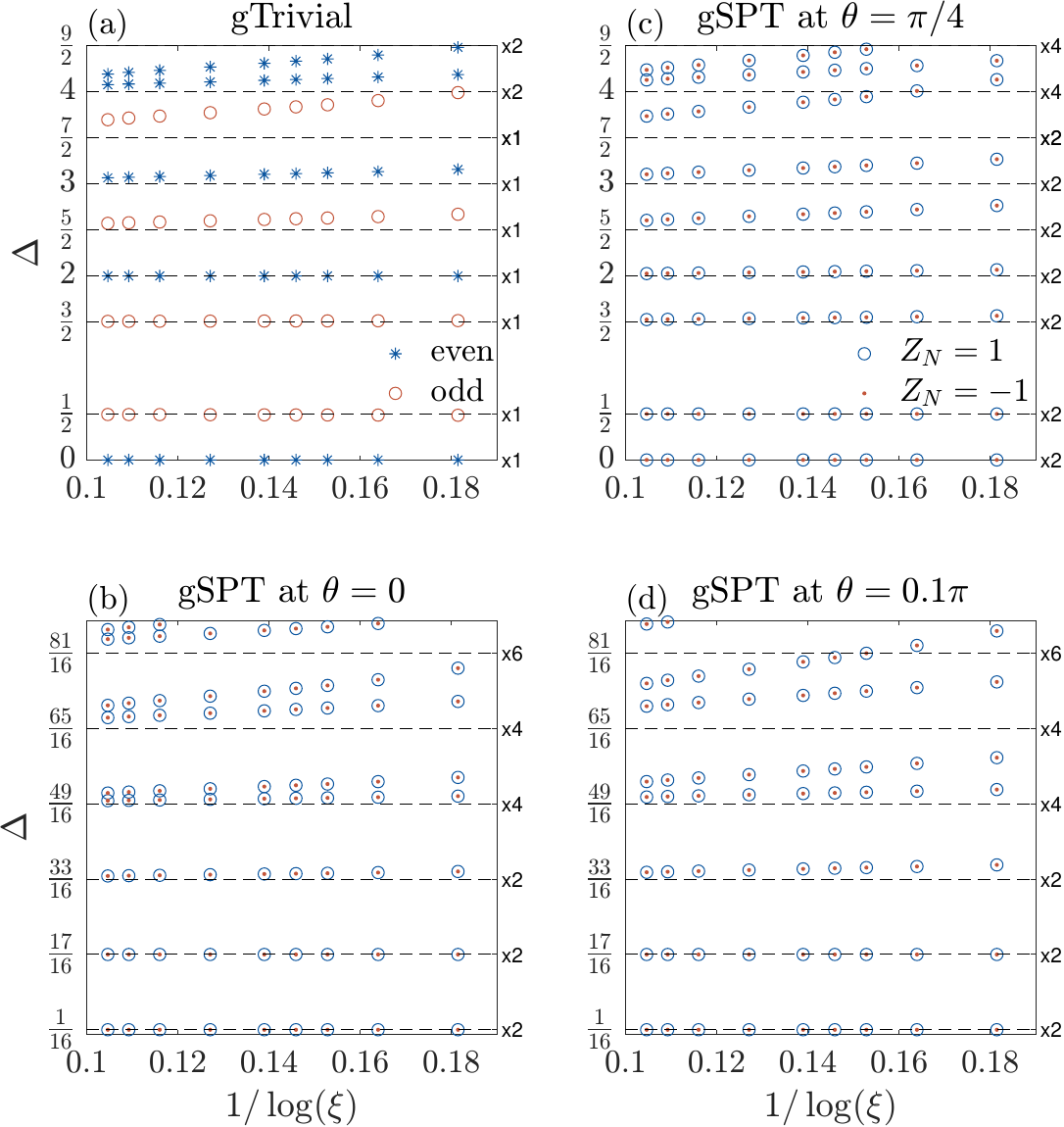}
    \caption{\textbf{Entanglement Spectrum (ES) of the $\mathbb{Z}_2\times\mathbb{Z}_2$ gSPT state with $c=1/2$.} 
The ES in (a) is extracted from the $\mathbb{Z}_2$-symmetric iMPS corresponding to free physical boundaries with bond dimensions $\chi$ ranging from $30$ to $250$, which approximate the ground state of the critical Ising chain. Here, $\xi$ denotes the correlation length induced by the finite-$\chi$ truncation. Since the iMPS is $\mathbb{Z}_2$-symmetric, the ES is separated into $\mathbb{Z}_2$ parity even (corresponding to the primary field $I$) and parity odd (corresponding to the primary field $\epsilon)$. The ES in (b), (c), and (d) are obtained by applying the quantum channel, Eq.~\eqref{eq:channel4CZX}, with different parametrizations $\theta$ to the iMPS in (a). Since $[\varrho_{\gSPT},Z_N]=0$ the degenerate levels are labeled by $Z_N=\pm1$.
The ES levels are rescaled by the entanglement gap and shifted, and labeled by $\Delta$ (see main text), enabling direct comparison with universal CFT predictions. 
(a) The ES of the trivial gSPT state $\ket{\Psi_{\gtri}}$.  
(b), (c), and (d) The ES of the non-trivial gapless SPT state $\ket{\Psi_{\gSPT}(\theta)}$ at $\theta=0$, $\theta=\frac{\pi}{4}$, and $\theta=\frac{\pi}{10}$, respectively.}
    \label{Fig:ES_Z2_Z2_c_half}
\end{figure}

From the Ising BCFT, it is known that there are three Cardy states $\ket{I}$, $\ket{\sigma}$, and $\ket{\epsilon}$~\cite{CARDY_1989}. The states $\ket{I}$ and $\ket{\epsilon}$ correspond to fixing the boundary spin of the Ising chain to up and down, respectively, while $\ket{\sigma}$ corresponds to a free boundary. Since both boundaries of the EH in Eq.~\eqref{eq:H_E_Ising} are free, the ES of $\ket{\Psi_{\gtri}}$ is determined by the fusion rule $\sigma\times\sigma = I + \epsilon$ and consists of the conformal towers of the primary fields $I$ and $\epsilon$. Their scaling dimensions are $\Delta_I = 0$ and $\Delta_\epsilon = 1/2$ leading to the towers $0,2,3,4,\ldots$ and $1/2, 3/2, 5/2, 7/2, \ldots$, respectively.  
The numerical results shown in Fig.~\ref{Fig:ES_Z2_Z2_c_half}a are consistent with this prediction.  Here we use infinite matrix product states (iMPS)  instead of finite MPS for convenience; the method for simulating different physical OBC using iMPS is described in Appendix~\ref{app:OBC_IMPS}. We further replace the finite-size scaling of finite systems with finite-entanglement scaling for iMPS~\cite{PhysRevB.78.024410, Pollmann_finite_ent_scaling_2009}, by replacing the system size $N$ with the iMPS correlation length $\xi$. Moreover, to directly compare with the CFT predictions $\{\Delta^{\text{CFT}}_i
\}$ (dashed lines in figures), the ES are rescaled and shifted as $\Delta_i=(\epsilon_i-\epsilon_0)(\Delta^{\text{CFT}}_1-\Delta^{\text{CFT}}_0
)/(\epsilon_1-\epsilon_0)+\Delta^{\text{CFT}}_0$, where $
\{\epsilon_i
\}$ are sorted entanglement energies.

We now turn to the ES of the non-trivial gSPT state. One might expect that the entanglement boundary condition is determined by the projective representation of the protecting symmetry. Since $\varrho_{\gSPT}(\theta)$ carries the same projective $\mathbb{Z}_2\times\mathbb{Z}_2$ representation for all $\theta$, this would suggest that the ES of $\ket{\Psi_{\gSPT}(\theta)}$ is described by the same BCFT. However, we find that this expectation is not correct in general.  To determine the entanglement boundary condition, we utilize the quantum channel. Since the representation of Kraus operators of a given quantum channel is not unique (they can be related by a unitary), we can rewrite the quantum channel in Eq.~\eqref{eq:channel4CZX} as
\begin{align}\label{eq:kraus_proj_Z2_Z2}
    \varrho_{\gSPT} &= \mathscr{N}_{CZ}[\mathscr{N}_X[\rho_{\gtri}]] \notag \\
    &= \sum_{j=0,1} \frac{1+(-1)^j Z_{N}}{2}\, \mathscr{N}_X[\rho_{\gtri}]\, \frac{1+(-1)^j Z_{N}}{2}.
\end{align}
In this representation, the Kraus terms are orthogonal and have an identical spectrum, leading to a two-fold degenerate ES.
When $\theta=0$, the qubit at site $N$, which is closest to the entanglement cut, is projected onto $\ket{0}$ or $\ket{1}$. When $\theta=\pi/4$, the qubit at site $N$ is instead projected onto $\ket{+_y}=e^{i\frac{\pi}{4}X}\ket{0}$ or $\ket{-_y}=e^{i\frac{\pi}{4}X}\ket{1}$, which are eigenstates of the Pauli $Y$ matrix. Taking into account the symmetry of the ES with respect to $\theta$, it is sufficient to restrict to $\theta\in[0,\pi/4]$.\footnote{From Eq.~\eqref{eq:channel4CZX}, the ES is $\pi/2$ periodic. This follows from $
\mathscr{N}_Z\!\left[e^{i(\pi/2+\theta)X_N}\rho_{\gtri}e^{-i(\pi/2+\theta)X_N}\right]
= X_N \mathscr{N}_Z\!\left[e^{i\theta X_N}\rho_{\gtri}e^{-i\theta X_N}\right] X_N
\cong \mathscr{N}_Z\!\left[e^{i\theta X_N}\rho_{\gtri}e^{-i\theta X_N}\right]$,
where we have used $e^{i(\pi/2+\theta)X}= i X e^{i\theta X}$ and the identity
$\mathscr{N}_Z[X_N\,\cdot\, X_N]=X_N\mathscr{N}_Z[\cdot]X_N$.  
Moreover, since $\rho_{\gSPT}(\theta)$ and $\rho_{\gSPT}(-\theta)$ are Hermitian and related by complex conjugation, they have identical ES. Therefore, the ES $\{\epsilon_i(\theta)\}$ satisfies both $\epsilon_i(\theta+\pi/2)=\epsilon_i(\theta)$ and $\epsilon_i(\theta)=\epsilon_i(-\theta)$. Any function with these properties is symmetric about $\theta=\pi/4$, i.e., $\epsilon_i(\pi/2-\theta)=\epsilon_i(\theta)$. Consequently, it is sufficient to consider the ES for $\theta\in[0,\pi/4]$.
}

The effective EH of the non-trivial gSPT state can be inferred by applying the Kraus projectors obtained from the quantum channel to $H_{E,\gtri}$, as illustrated in Fig.~\ref{fig:bulk_boundary}b. When $\theta=0$, the qubit at site $1$ of $H_{E,\gtri}$ in Eq.~\eqref{eq:H_E_Ising} is projected onto $\ket{0}$ and $\ket{1}$, yielding the EH of the non-trivial gSPT state:
\begin{align}\label{eq:H_E_Ising_mix}
    H_{E,\gSPT}(0) &\sim -\sum_{n=1}^{L-1} Z_{n} Z_{n+1} - \sum_{n=2}^{L} X_{n}.
\end{align}
Compared to $H_{E,\gtri}$, the entanglement boundary condition is changed from free to mixed, such that $[H_{E,\gSPT}(0), Z_1]=0$. The sector with $Z_1=1$ ($Z_1=-1$) corresponds to the Cardy state $\ket{I}$ ($\ket{\epsilon}$). Using the fusion rule $(I+\epsilon)\times\sigma=\sigma+\sigma$ and the fact that the scaling dimension is $\Delta_{\sigma}=1/16$, the ES of $\ket{\Psi_{\gSPT}(0)}$ consists of two copies of the conformal tower $1/16,17/16,33/16,\cdots$ associated with the primary field $\sigma$ (irrespective of $Z_1=\pm 1$).  
To numerically verify this analysis, we obtain the ES of $\ket{\Psi_\gSPT(\theta)}$ by directly applying the quantum channel to the optimized iMPS of the Ising chain, see details in Appendix~\ref{app:OBC_IMPS}. The numerical results shown in Fig.~\ref{Fig:ES_Z2_Z2_c_half}b are in good agreement with the above analysis and are consistent with those obtained directly from the non-trivial SPT states~\cite{Huang_2024}.

When $\theta=\pi/4$, to obtain the effective EH of the non-trivial gSPT state, we project the qubit at site $1$ of $H_{E,\gtri}$ in Eq.~\eqref{eq:H_E_Ising} onto $\ket{\pm_y}$, which yields the EH of the non-trivial gSPT state:
\begin{align}\label{eq:EH_pi_quarter}
    H_{E,\gSPT}\!\left(\frac{\pi}{4}\right) &\sim -\sum_{n=2}^{L-1} Z_{n} Z_{n+1} - \sum_{n=2}^{L} X_{n}.
\end{align}
Note that the entanglement boundary condition remains free; however, the EH no longer acts on site $1$, which leads to a two-fold degeneracy. Consequently, the ES of $\ket{\Psi_{\gSPT}\!\left(\frac{\pi}{4}\right)}$ consists of two copies of the conformal towers associated with the primary fields $I$ and $\epsilon$. The numerical results shown in Fig.~\ref{Fig:ES_Z2_Z2_c_half}c are consistent with this prediction.

\begin{figure}
     \centering
    \includegraphics[width=0.75\linewidth]{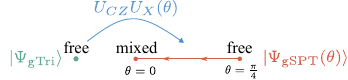}
    \caption{\textbf{The RG flow of the entanglement boundary.} 
A family of non-trivial gSPT states $\ket{\Psi_{\gSPT}(\theta)}$ within the same $\mathbb{Z}_2\times\mathbb{Z}_2$ gSPT phase is constructed by the $\theta$-parameterized unitary SPT entanglers $U_{CZ}U_X(\theta)$. The ES at $\theta=0$ and $\theta=\frac{\pi}{4}$ are described by different BCFT of the same CFT. There is a boundary RG flow of the entanglement boundary from the free boundary condition at $\theta=\pi/4$ to the mixed boundary condition $\theta=0$, driven by the fact that the boundary entropy at $\theta=0$ is lower than the one at $\theta=\pi/4$.}
    \label{fig:RG}
\end{figure}

What happens when $\theta\in(0,\pi/4)$? This question can be addressed using boundary renormalization (RG) flow. We consider the boundary entropy of a given entanglement boundary condition. For a given CFT, each conformal boundary condition is characterized by a universal Affleck--Ludwig boundary entropy, which is a non-increasing function along the boundary RG flow~\cite{Affleck_Ludwig_1991,CARDY_2006}. For the Ising CFT, the free and fixed boundary conditions have boundary entropies $S_{\free}=0$ and $S_{\fix}=-\ln(2)/2$, respectively~\cite{Affleck_Ludwig_1991}. Therefore, the boundary entropy difference between the entanglement boundaries of the non-trivial gSPT states at $\theta=0$ and $\theta=\pi/4$ is
$S_{\fix}-S_{\free}=-\ln(2)/2$. Since the boundary entropy at $\theta=0$ is smaller than that at $\theta=\pi/4$, the ES at $\theta=0$ is more stable than that at $\theta=\pi/4$. Consequently, for $\theta\in(0,\pi/4)$, the ES flows under boundary RG to the BCFT describing the ES at $\theta=0$, as schematically illustrated in Fig.~\ref{fig:RG}. This scenario is further supported by the numerical ES of $\ket{\Psi_{\gSPT}(\frac{\pi}{10})}$ shown in Fig.~\ref{Fig:ES_Z2_Z2_c_half}d.

\begin{figure*}
     \centering
    \includegraphics[width=\linewidth]{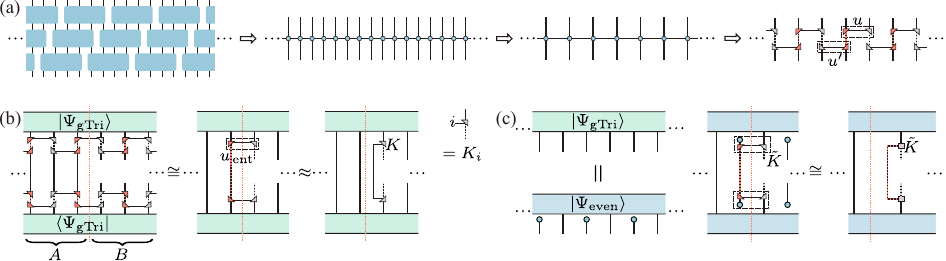}
    \caption{\textbf{The framework.} (a) An SPT entangler, which is typically a finite-depth quantum circuit, can be represented as an MPU. A simple MPU tensor is obtained by blocking MPU tensors. By decomposing the simple MPU tensors, the MPU can be brought into a standard form consisting of unitary gates $u$ and $u'$. 
(b) The reduced density matrix of the non-trivial gSPT state satisfies
$\rho_{\gSPT}=\Tr_A\!\left(U\rho_{\gtri}U^{\dagger}\right)\cong\Tr_A\!\left(u_{\ent}\rho_{\gtri}u_{\ent}^{\dagger}\right)\approx \sum_i K_{i}\rho_{\gtri}K_{i}^\dagger$,
where $u_{\ent}$ is the gate crossing the entanglement cut and $K_i$ are the corresponding Kraus operators.
(c) When the trivial gSPT state is a tensor product of a product state (the blue dots) on the odd sites and a critical state on the even site, the quantum-channel relation becomes exact, and $\tilde{K}$ and $K$ are equivalent up to a unitary transformation in the situations considered in this work.}
    \label{fig:framework}
\end{figure*}

In fact, at $\theta=0$ and $\theta=\pi/4$, the non-trivial gSPT states possess not only the $\mathbb{Z}_2\times\mathbb{Z}_2$ symmetry but also an additional $\mathbb{Z}_2\times\mathbb{Z}_2^T$ symmetry, where $\mathbb{Z}_2$ is generated by the global spin flip $\prod_{n=1}^{2N-1} X_n$, and $\mathbb{Z}_2^T$ denotes time-reversal symmetry $T$, implemented as complex conjugation.  In Sec.~\ref{Sec:Z_2T_and_c_half} and Appendix~\ref{app:same_phase}, we show that $\ket{\Psi_{\gSPT}(0)}$ and $\ket{\Psi_{\gSPT}\!\left(\frac{\pi}{4}\right)}$ are related to critical spinless free-fermion chains with time-reversal symmetry in the BDI class, whose topological invariants are $\omega=1$ and $\omega=2$, respectively~\cite{Ruben_2018}. Consequently, the ES at $\theta=0$ and $\theta=\pi/4$ can be different.

\section{General framework}\label{sec:framework}

We now extend the analysis of the ES for $\mathbb{Z}_2\times\mathbb{Z}_2$ gSPT states to a broader class of non-trivial gSPT states obtained by applying SPT entanglers $U$ to critical states~\cite{Scaffidi_gSPT_2017,Ruben_SPT_and_transition_2017}, i.e., $\ket{\Psi_{\gSPT}} = U \ket{\Psi_{\gtri}}$, as illustrated in Fig.~\ref{fig:problem}c. In particular, we introduce a general framework to analyze and predict the ES of non-trivial gSPT states $\ket{\Psi_{\gSPT}}$.

Our framework is based on relating the gSPT states via SPT entanglers. Hence, we consider the relation between the reduced density matrices $
\rho_{\gtri} = \Tr_A \ket{\Psi_{\gtri}}\bra{\Psi_{\gtri}}$ and $
\rho_{\gSPT} = \Tr_A \ket{\Psi_{\gSPT}}\bra{\Psi_{\gSPT}}
$. For simplicity, we focus on OBC chains with a single entanglement cut in the middle. Generally, SPT entanglers $U$ can be expressed as matrix product unitaries (MPU) associated with some non-trivial projective representations of the protecting symmetry group~\cite{MPU_Chen_2018,MPU_cirac_2017,MPU_gong_2020}. As illustrated in Fig.~\ref{fig:framework}a, to reveal some fixed point properties of MPU, one needs to block sufficiently many MPU tensors to obtain a new MPU tensor referred to as simple MPU tensor. By applying singular value decomposition to a simple MPU tensor, the MPU can be brought into a standard form with unitary gates $u$ and $u'$~\cite{MPU_cirac_2017,MPU_Chen_2018}. Only the gate $u_{\ent}$ crossing the entanglement cut affects the ES, while all other gates can be ignored. Therefore,
\begin{equation}
    \rho_{\gSPT} = \Tr_A\left(U \rho_{\gtri} U^{\dagger}\right) \cong \Tr_A\left(u_{\ent} \rho_{\gtri} u_{\ent}^{\dagger}\right),
\end{equation}
as shown in Fig.~\ref{fig:framework}b, where ``$\cong$'' indicates that the two reduced density matrices have identical ES.\footnote{The MPU admits two standard forms. In one form, a blocked MPU tensor on an even (odd) site is decomposed into two red (gray) triangular tensors, as shown in Fig.~\ref{fig:framework}. In the other form, a blocked MPU tensor on an even (odd) site is decomposed into two gray (red) triangular tensors. When the entanglement cut is shifted by one site, the alternative standard form can be used.} Using the properties of the MPU, we derive that the relation between the reduced density matrices of the trivial and non-trivial gSPT states takes the form
\begin{align}\label{eq:channel}
    \rho_{\gSPT} \cong \mathscr{N}[\rho_{\gtri}] + \varsigma = \varrho_{\gSPT} + \varsigma.
\end{align}
Here, $\varrho_{\gSPT}$ is obtained by applying a quantum channel $\mathscr{N}[\cdot]$ to $\rho_{\gtri}$: 
\begin{equation}\label{eq:quantumchannel}
    \varrho_{\gSPT} = \mathscr{N}[\rho_{\gtri}] = \sum_i K_{i} \rho_{\gtri} K_{i}^{\dagger},
\end{equation}
where $K_{i}$ are the Kraus operators acting on the degrees of freedom in the vicinity of the entanglement cut, see Fig.~\ref{fig:framework}b. The operator $\varsigma$ is traceless and Hermitian. The explicit forms of $K_{i}$ and $\varsigma$ depend on the MPU, and a detailed derivation is provided in Appendix~\ref{app:channel}.

When $\ket{\Psi_{\gtri}}$ is a tensor product of a critical state and a product state, as shown in the illustrative example of Sec.~\ref{sec:c_half_z2_z2} and Fig.~\ref{fig:framework}c, we find $\varsigma = 0$,  and the ES obtained from $\rho_{\gSPT}$ and $\varrho_{\gSPT}$ are exactly identical. This scenario also arises in the context of gapped 1-form SPT states in (2+1)D lattice gauge theories~\cite{xu_2024_entanglement}.  
In general, $\varsigma \neq 0$ and its consequence needs to be analyzed separately. However, for all the different classes of gSPT states studied in this work, we are able to show that $\varsigma$ does not alter the universal properties of the ES.
We will investigate
the reason below, and now explain how to predict the entanglement boundary condition using the quantum channel relation, Eq.~\eqref{eq:quantumchannel}, ignoring the contribution of $\varsigma$. 
For a given quantum channel, the Kraus operators are not unique: one can define new Kraus operators as linear combinations of the original ones, $P_{i}=\sum_j W_{ij} K_{j}$, where $W$ is an arbitrary unitary matrix.  In the examples considered in this work, one can find a unitary matrix $W$ such that the resulting Kraus operators $P_{i}$ are projectors: $P^2_{i} \propto P_{i}$. 

Since the projectors $P_{i}$ fix the degrees of freedom of $\rho_{\gtri}$ near the entanglement cut, the entanglement boundary condition may be changed from free to fixed. When summing over all projectors, the entanglement boundary condition is a mixture of different fixed boundary conditions, i.e., a mixed boundary condition. This explains, via the quantum channel relation, how the MPU modifies the entanglement boundary condition.  
In practice, the entanglement boundaries can be more complicated, so it is useful to analyze the effective EH. Given the effective EH of the trivial gSPT state, $H_{E,\gtri} \sim -\ln(\rho_{\gtri})$, the effective EH corresponding to $\varrho_{\gSPT}$ can be estimated using $H_{E,\gSPT} \sim -\ln \left( \mathscr{N} \left[ \exp(-H_{E,\gtri}) \right]\right)$.
In order to obtain the universal low-energy part of the ES, it is sufficient to consider only the leading terms, i.e., $\mathscr{N}[H_{E,\gtri}]$, as shown in Fig.~\ref{fig:bulk_boundary}b. 

Once the entanglement boundary condition is determined from the effective EH, we analyze the ES using the framework of BCFT.
In a BCFT, the partition function on a cylinder is specified by two conformal boundaries, each characterized by a Cardy state. Cardy states are in one-to-one correspondence with the primary fields of the underlying CFT~\cite{CARDY_1984,CARDY_1986,CARDY_1989,CARDY_2006}. Accordingly, we label the conformal boundaries by the primary field labels $a$ and $b$. Given the conformal boundaries $a$ and $b$, the ES is described by the conformal towers associated with the primary fields $d$ appearing in the fusion rule
$a\times b=\sum_d N_{ab}^d\, d$~\cite{Cho_ES_BCFT_PRB_2017},
where $N_{ab}^d$ denotes the fusion tensor. As a consequence of the fusion rules, the ES may be completely different after applying the MPU corresponding to distinct primary fields of the BCFT.

We now analyze under which conditions the term $\varsigma$ in Eq.~\eqref{eq:channel} can be neglected. Non-intrinsic gSPT phases separate into two classes: those with gapped sectors~\cite{Scaffidi_gSPT_2017} and those without gapped sectors~\cite{Ruben_SPT_and_transition_2017,Gapless_SPT_Ruben_2021}. We discuss these two classes separately. 

Case (I): The illustrative example of Sec.~\ref{sec:c_half_z2_z2} shows gSPT states with gapped sectors. Such states can exhibit exponentially decaying correlation functions. They are typically obtained from the decorated domain wall construction~\cite{Chen_decorated_DW_2014} and are closely related to Higgs–deconfined phase transitions in one-dimensional lattice gauge theories~\cite{Gauging_Kitaev_2021,Gapless_SPT_Ruben_2021,1_form_QEC_2025}. When the gapped sector of $\ket{\Psi_{\gtri}}$ is not a product state, one can perform RG transformations~\cite{MPS_RG_2005} that preserves the protecting symmetry, such that the gapped sector flows to a symmetric product state and $\ket{\Psi_{\gtri}}$ becomes a tensor product of a critical state and a product state. Since RG transformations do not change the gSPT phase, the ES before and after the RG flow share the same universal properties, except in cases where the symmetry is accidentally enlarged at fine tuned points (\textit{c.f.} non-trivial gSPT states for $\theta=0$ and $\pi/4$ in Sec.~\ref{sec:c_half_z2_z2}). Therefore, the contribution of $\varsigma$ are irrelevant for gSPT states with gapped sectors.

\begin{figure}
     \centering
\includegraphics[width=1.0\linewidth]{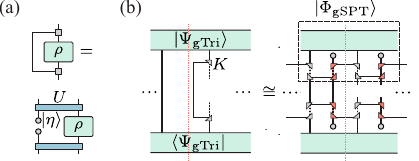}
    \caption{\textbf{Application of Stinespring’s dilation to gSPT states without gapped sectors.}
(a) Graphical representation of Stinespring’s dilation theorem. The quantum channel is represented by Kraus operators denoted as the gray boxes.
(b) Application of the theorem to the relation $\varrho_{\gSPT}=\mathscr{N}(\rho_{\gtri})$. By introducing an ancillary product state $\ket{\eta}=\ket{\pmb{1}}_{\text{ac}}$ (gray circles) and identifying the MPU with the unitary in the dilation, we construct a state $\ket{\Phi_{\gSPT}}$ that contains a gapped sector and whose reduced density matrix is equivalent to $\varrho_{\gSPT}$.} 
    \label{Fig:Stinespring_Dilation}
\end{figure}
Case (II): The situation is more involved when considering gSPT states without gapped sectors. We can make progress by invoking Stinespring’s dilation theorem~\cite{stinespring_1955} from quantum information theory. This theorem states that any quantum channel acting on a density matrix $\rho$ can be written as
$
\mathscr{N}(\rho)
= \operatorname{tr}_A \!\left[
U \left( \rho \otimes \ket{\eta}\bra{\eta} \right) U^\dagger
\right],
$
where $\ket{\eta}$ is a pure state of an ancillary system and $U$ is a unitary operator, as illustrated in Fig.~\ref{Fig:Stinespring_Dilation}a.
Applying this theorem to the relation $\varrho_{\gSPT} = \mathscr{N}(\rho_{\gtri})$, we introduce a symmetric product state $\ket{\pmb{1}}_{\text{ac}}$ as the ancillary system and identify the MPU with the unitary operator $U$. This allows us to define a state
$
\ket{\Phi_{\gSPT}}
= U \bigl( \ket{\Psi_{\gtri}} \otimes \ket{\pmb{1}}_{\text{ac}} \bigr)
$ containing a gapped sector, and it reduced density matrix is exactly equivalent to $\varrho_{\gSPT}$ in the sense that they possess identical ES, as illustrated in Fig.~\ref{Fig:Stinespring_Dilation}b. 
The key point of the argument is: if one can show that $\ket{\Phi_{\gSPT}}$ and $\ket{\Psi_{\gSPT}}\otimes \ket{\pmb{1}}_{\text{ac}}$ ($\ket{\pmb{1}}_{\text{ac}}$ does not affect ES) are smoothly connected by a symmetry-preserving path, then the ES derived from $\varrho_{\gSPT}$ and $\rho_{\gSPT}$ share the same universal properties. 
Because the boundary entropy is a non-increasing function along boundary RG flows~\cite{Affleck_Ludwig_1991,CARDY_2006}, the entanglement boundary condition flows, among all those allowed by the projective symmetry, to the conformal boundary that has the lowest boundary entropy~\cite{Roy_ent_boundary_2025}. 
One can show that $\rho_{\gSPT}$ and $\varrho_{\gSPT}$ carry exactly the same projective symmetries inherited from the unitary protecting symmetry of the gSPT phase; see Appendix~\ref{app:channel}. Therefore, $\varsigma$ can be regarded as a symmetric perturbation that deforms the EH obtained from $\varrho_{\gSPT}$ to that from $\rho_{\gSPT}$.
As long as this perturbation does not introduce any additional symmetry into $\rho_{\gSPT}$, it is irrelevant. Consequently, the universal properties of the ES extracted from $\rho_{\gSPT}$ and $\varrho_{\gSPT}$ are identical, since the BCFT describing the low-energy ES is universal and independent of microscopic details.

Concrete examples illustrating the existence of an adiabatic and symmetric path connecting $\ket{\Psi_{\gSPT}}\otimes \ket{\pmb{1}}_{\text{ac}}$ and $\ket{\Phi_{\gSPT}}$ are presented in the following section. Crucially though, it seems that $\ket{\Psi_{\gSPT}}\otimes \ket{\pmb{1}}_{\text{ac}}$ and $\ket{\Phi_{\gSPT}}$ do not have to belong to the same gSPT phase and if so, $\varsigma$ can qualitatively change the universal properties of the ES. Therefore, it remains an interesting direction to analyze under which conditions adiabatic paths connecting $\ket{\Psi_{\gSPT}}\otimes \ket{\pmb{1}}_{\text{ac}}$ and $\ket{\Phi_{\gSPT}}$ can be constructed.

We conclude with several remarks on the framework, which helps to systematically analyze gSPT states:

\noindent (i) Knowing only the projective representation of the protecting symmetry acting on $\rho_{\gSPT}$ is generally insufficient to uniquely determine the conformal boundary condition of the entanglement boundary, since a gSPT state may possess additional symmetries that are not manifest.  In such cases, the quantum channel provides a direct and systematic way to extract the effective EH without requiring complete knowledge of all symmetries of the gSPT state, as we demonstrate in Sections~\ref{sec:c_half_z2_z2} and~\ref{Sec:c=1_Z2_Z2}. A similar issue arises in gapped SPT phases, where additional symmetries may emerge at fine-tuned points within the phase, leading to extra accidental degeneracies.

\noindent (ii) Because our framework is based on MPUs and does not rely on the specific type of protecting symmetry, it can be applied to non-invertible matrix product operator symmetry-protected SPT states that are constructed using MPUs~\cite{Nat_2025}. We discuss such examples in Sec.~\ref{sec:non_inv_SPT}.

\noindent (iii) The framework can be extended to analyze the ES of gSPT states with periodic boundary conditions, where two entanglement cuts are present. Once the Cardy states associated with each entanglement cut are identified using the quantum channel, the ES follows directly from the fusion rule of the two corresponding Cardy states.

\noindent (iv) The quantum channel is a non-invertible transformation due to the presence of zero eigenvalues. In the example of the $\mathbb{Z}_2\times\mathbb{Z}_2$ gSPT states, although applying the MPU twice maps the system back to the trivial state, one cannot recover the reduced density matrix of the trivial state by applying the quantum channel twice. 

\noindent (v) An important quantity characterizing an MPU is its index, which measures the net flow of information through a cut of the MPU. In Appendix~\ref{app:MPU_of_translation}, we analyze $\mathbb{Z}_2\times\mathbb{Z}_2$ gSPT states with $c=1/2$ that are distinct from those in Sec.~\ref{sec:c_half_z2_z2}  
and are related by MPUs with nonzero index. In these cases, we show that the corresponding reduced density matrices are always exactly related by a quantum channel and our framework is also valid when the MPU has a non-zero index.

In the following, we apply our framework to additional representative examples and demonstrate its generality.

\section{ES of $\mathbb{Z}_2\times\mathbb{Z}_2^T$ gSPT states with central charge $c=\frac{1}{2}$}\label{Sec:Z_2T_and_c_half}

In Sec.~\ref{sec:c_half_z2_z2}, we demonstrated that the framework works well for gSPT states with gapped sectors. In this section, we discuss pure gSPT states without gapped sectors to demonstrate that the framework is also applicable in this case.

\begin{table}
\centering
\begin{tabular}{c|c}
\hline
$\alpha$ & $H_{\alpha}$ \\
\hline
$-3$ & $-\sum_n Y_nX_{n+1}X_{n+2}Y_{n+3}$ \\
$-2$ & $-\sum_n Y_nX_{n+1}Y_{n+2}$ \\
$-1$ & $-\sum_n Y_nY_{n+1}$ \\
$0$  & $-\sum_n X_n$ \\
$1$  & $-\sum_n Z_nZ_{n+1}$ \\
$2$  & $-\sum_n Z_nX_{n+1}Z_{n+2}$ \\
$3$  & $-\sum_n Z_nX_{n+1}X_{n+2}Z_{n+3}$ \\
$4$  & $-\sum_n Z_nX_{n+1}X_{n+2}X_{n+3}Z_{n+4}$ \\
\hline
\end{tabular}
\caption{The eight fixed points of 1D gapped systems with $\mathbb{Z}_2\times\mathbb{Z}_2^T$ symmetry are given by the Hamiltonians $H_\alpha$.}
\label{tab:alpha_chain}
\end{table}

A class of gSPT states without gapped sectors can be constructed from the so-called $\alpha$-chains~\cite{Ruben_SPT_and_transition_2017,Ruben_2018}, as shown in Tab.~\ref{tab:alpha_chain}. These models have the $\mathbb{Z}_2\times\mathbb{Z}_2^T$ symmetry generated by global spin flip and complex conjugation, and they serve as fixed-point models of gapped SPT phases. Since the classification of 1D gapped $\mathbb{Z}_2\times\mathbb{Z}_2^T$ systems is $\mathbb{Z}_8$~\cite{Turner-Pollmann-Berg_2011,Fidkowski-Kitaev-2011}, we consider eight $\alpha$-chains in the spin representation.  
When $\alpha$ is even, the chains are fixed-point models for four different gapped $\mathbb{Z}_2\times \mathbb{Z}_2^T$ SPT phases. When $\alpha$ is odd, they are fixed-point models that spontaneously break the $\mathbb{Z}_2$ spin-flip symmetry. A $\mathbb{Z}_2\times\mathbb{Z}_2^T$ gSPT model can be constructed as $H_{\alpha,\alpha'}=H_{\alpha}+H_{\alpha'}$, which has a central charge $c=\frac{|\alpha-\alpha'|}{2}$~\cite{Ruben_SPT_and_transition_2017}. For simplicity, we focus on the case $c=\frac{1}{2}$, leading to seven models. Since $H_{\alpha}$ and $H_{-\alpha}$ are related by the on-site unitary $U_X(\frac{\pi}{4})$, the pairs $H_{\alpha,\alpha+1}$ and $H_{-\alpha,-(\alpha+1)}$ have the same ES, so it suffices to consider the four models $H_{0,1}, H_{1,2}, H_{2,3}$, and $H_{3,4}$.  

In the Majorana representation, the $\alpha$-chains are spinless free fermion chains with $\mathbb{Z}_2^T$ symmetry, and $H_{\alpha,\alpha+1}$ with distinct $\alpha$ are characterized by the BDI topological invariant $\omega=\alpha$~\cite{Ruben_2018}, confirming that they realize distinct $\mathbb{Z}_2\times\mathbb{Z}_2^T$ gSPT states. Their ES has been studied in the free fermion representation~\cite{Yu_2025,Meng_2025}. Here, we focus on the spin representation and identify the BCFT describing the ES.  
Among these four models, $H_{1,2}, H_{2,3}$, and $H_{3,4}$ are obtained from $H_{0,1}$ by applying appropriate MPUs~\cite{Pivot_2023}:
\begin{align}\label{eq:MPUs_and_alpha_chain}
   H_{1,2}&=U_{1}H_{0,1}U^{\dagger}_{1}, \quad U_1=U_{CZ},\notag\\ 
   H_{2,3}&= U_2H_{0,1} U_2^{\dagger},\quad U_2=U_{CZ}U_X\big(\frac{\pi}{4}\big),\notag\\  
   H_{3,4}&=U_3H_{0,1}U_3^{\dagger},\quad U_3=U_{CZ}U_X\big(\frac{\pi}{4}\big)U_{CZ}.
\end{align}  
Here, $H_{0,1}$ is simply the transverse-field Ising chain and has no gapped sector. Since the MPU preserves locality all four models are gapless SPT states without gapped sectors. We identify $H_{0,1}$ as the reference model from which the other three are obtained via MPU transformations. Therefore, the ES of all four models can be extracted from the ground state of $H_{0,1}$, the critical OBC Ising chain with free physical boundaries. Furthermore, as the ground state $\ket{\Psi_{\Ising}}$ of $H_{0,1}$ differs from that of the model in Eq.~\eqref{eq:Z2_Z2_c_half_gtri} only by a product state, their ES are identical and Fig.~\ref{Fig:ES_Z2_Z2_c_half}a also represents the ES of $H_{0,1}$.

The ground state of the model $H_{\alpha,\alpha+1}$ is $\ket{\Psi_{\alpha,\alpha+1}}=U_{\alpha}\ket{\Psi_{\Ising}}$. Here $\ket{\Psi_{1,2}}$ and $\ket{\Psi_{2,3}}$ are very similar to $\ket{\Psi_{\gSPT}(0)}$ and $\ket{\Psi_{\gSPT}(\frac{\pi}{4})}$ in Eq.~\eqref{eq:psi_gSPT}, respectively, with the main difference being that the latter two contain gapped sectors. In fact, $\ket{+}_{\odd}\otimes\ket{\Psi_{1,2}}_{\even}$ and $\ket{\Psi_{\gSPT}(0)}$ can be smoothly connected along a $\mathbb{Z}_2\times\mathbb{Z}_2^T$ symmetric path, and similar for $\ket{+}_{\odd}\otimes\ket{\Psi_{3,4}}_{\even}$ and $\ket{\Psi_{\gSPT}(\frac{\pi}{4})}$, see Appendix~\ref{app:same_phase}, indicating that they belong to the same $\mathbb{Z}_2\times\mathbb{Z}_2^T$ gSPT phase. 

\begin{figure}
    \centering
        \includegraphics[scale=0.5]{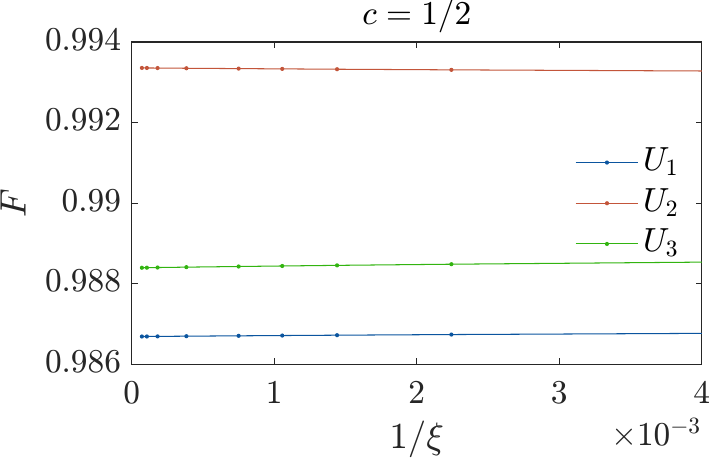}
    \caption{\textbf{Fidelity between reduced density matrices for non-trivial $\mathbb{Z}_2 \times \mathbb{Z}_2^T$ gSPT states with $c =1/2$.} The Fidelity is defined as $F=\Tr\!\Big(\sqrt{\sqrt{\varrho_{\alpha,\alpha+1}}\,(\varrho_{\alpha,\alpha+1}+\varsigma_{\alpha,\alpha+1})\,\sqrt{\varrho_{\alpha,\alpha+1}}}\Big)$. The results are obtained by applying the MPU $U_{\alpha}$ to the $\mathbb{Z}_2$-symmetric iMPS ($\chi = 30$–$250$) representing the ground state of the critical Ising chain.}
    \label{Fig:Fidelity_c_half}
\end{figure}

The ground state reduced density matrices $\rho_{\alpha,\alpha+1}$ and $\rho_{\Ising}$ are related by: 
\begin{align}\label{eq:not_exact_channel_relation}
    \rho_{\alpha,\alpha+1}&\cong\mathscr{N}_{\alpha}[\rho_{\Ising}]+\varsigma_{\alpha,\alpha+1}=\varrho_{\alpha,\alpha+1}+\varsigma_{\alpha,\alpha+1},
\end{align}
where $\varsigma_{\alpha,\alpha+1}$ is the traceless Hermitian operator introduced in Eq.~\eqref{eq:channel} and
\begin{align}\label{eq:def_3_channel}
    \mathscr{N}_{1}[\cdot]&=\mathscr{N}_{CZ}[\cdot], \quad \mathscr{N}_{2}[\cdot]=\mathscr{N}_{CZ}[\mathscr{N}_X[\cdot]]\notag\\
    \mathscr{N}_{3}[\cdot] &= \sum_{i=0,1} \frac{1 + (-1)^i X_{N} Z_{N+1}}{2} \, \mathscr{N}_{CZ}[\cdot] \, \frac{1 + (-1)^i X_{N} Z_{N+1}}{2},
\end{align}
with $\mathscr{N}_{CZ}[\cdot]$ and $\mathscr{N}_{X}[\cdot]$ being defined in Eq.~\eqref{eq:channel4CZX}. By applying the Stinespring's dilation to the quantum channels, we have $\varrho_{1,2}\cong\varrho_{\gSPT}(0)$ and $\varrho_{3,4}\cong\varrho_{\gSPT}(\frac{\pi}{4})$, see Sec.~\ref{sec:framework} and Appendix~\ref{app:same_phase}. In practice, the fidelity between $\varrho_{\alpha,\alpha+1}+\varsigma_{\alpha,\alpha+1}$ and $\varrho_{\alpha,\alpha+1}$ is very close to $1$, see Fig.~\ref{Fig:Fidelity_c_half}, indicating that $\varsigma_{\alpha,\alpha+1}$ can be viewed as a small symmetric perturbation. Therefore, it is expected that $\varsigma_{\alpha,\alpha+1}$ is irrelevant and can be ignored when analyzing the ES.

\begin{figure}
    \centering
        \includegraphics[width=\linewidth]{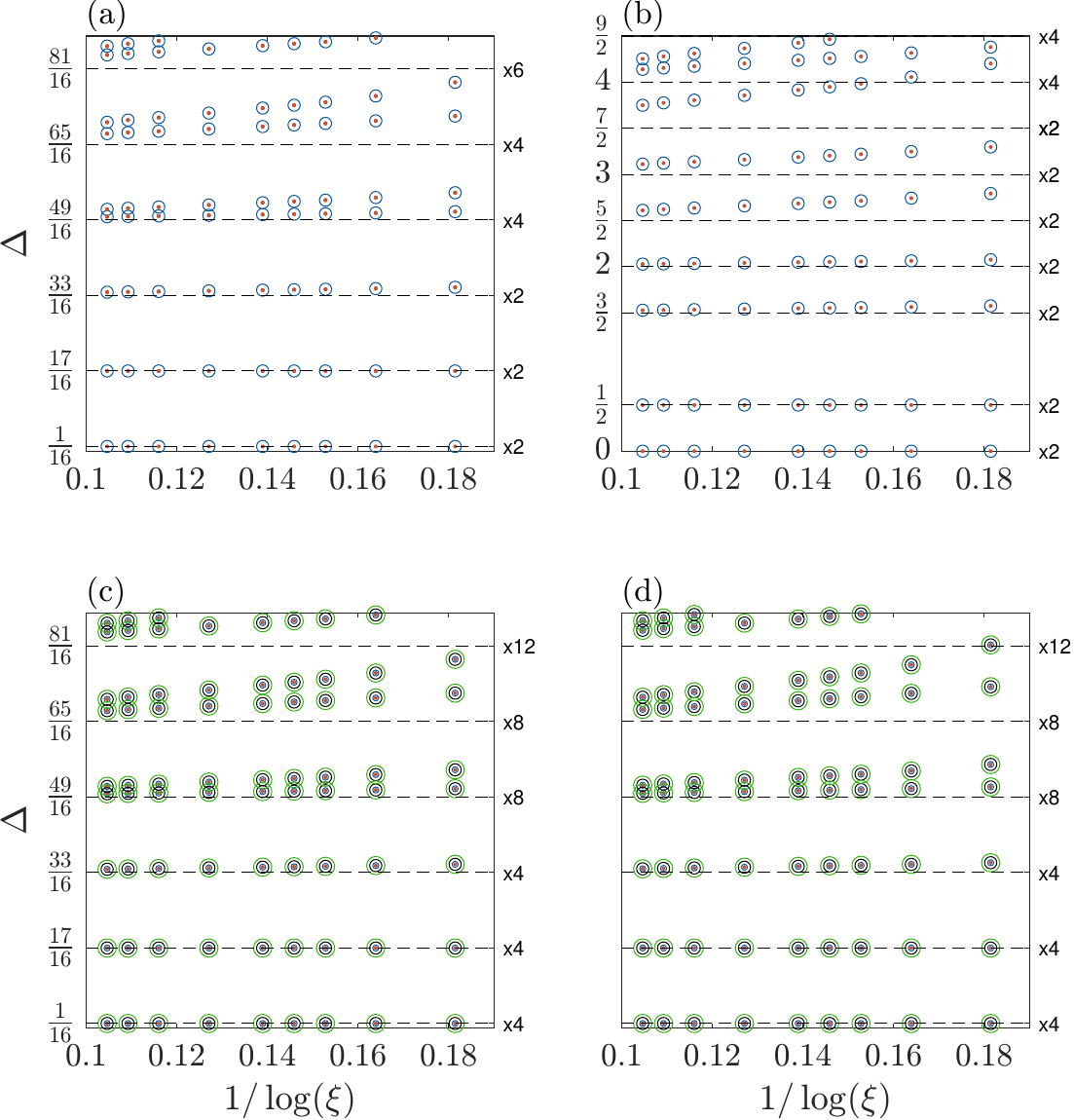}
    \caption{\textbf{Entanglement Spectra (ES) of the $\mathbb{Z}_2 \times \mathbb{Z}_2^T$ gSPT states with $c=1/2$.} 
All ES are extracted from the $\mathbb{Z}_2$-symmetric iMPS (corresponding to free physical boundaries) of the critical Ising chain with bond dimensions $\chi = 30$–$250$. To compare directly with the CFT predictions (dashed lines), the ES have been appropriately shifted and rescaled. 
(a) ES of $\ket{\Psi_{1,2}}$. 
(b) ES of $\ket{\Psi_{2,3}}$. 
(c) ES from $\varrho_{3,4} = \mathscr{N}_{3}[\rho_{\Ising}]$, see Eq.~\eqref{eq:not_exact_channel_relation}. 
(d) ES of $\ket{\Psi_{3,4}}$.}
    \label{Fig:ES_Ising_Z2_T}
\end{figure}

Indeed we find that the ES obtained from $\rho_{\alpha,\alpha+1}$ exhibits the same universal properties as the ES from $\varrho_{\alpha,\alpha+1}$. This can be seen by comparing the ES of $\rho_{1,2}$ shown in Fig.~\ref{Fig:ES_Ising_Z2_T}a with the ES of $\varrho_{1,2}$ shown in Fig.~\ref{Fig:ES_Z2_Z2_c_half}b, and the ES of $\rho_{2,3}$ in Fig.~\ref{Fig:ES_Ising_Z2_T}b with the ES of $\varrho_{2,3}$ in Fig.~\ref{Fig:ES_Z2_Z2_c_half}c. 
The minor differences arise from small non-universal finite-size effects, expected from our theoretical analysis. We also note that the OBC energy spectrum and the PBC ground state ES of $H_{1,2}$ have been studied previously~\cite{Yu_huang_2022,Universal_ES_2024}, and the entanglement boundary identified there is consistent with our predictions. 

By ignoring $\varsigma_{\alpha,\alpha+1}$ in Eq.~\eqref{eq:not_exact_channel_relation}, the effective EH of $\rho_{1,2}$ is Eq.~\eqref{eq:H_E_Ising_mix} and the effective EH of $\rho_{2,3}$ is Eq.~\eqref{eq:EH_pi_quarter}. The effective EH of $\rho_{3,4}$ can be considered in the same way. Since the effective EH $H_{E,\gtri}$ of $\rho_{\Ising}$ corresponds to the Ising chain with free boundary condition at both ends, see Eq.~\eqref{eq:H_E_Ising}, the effective EH of $\varrho_{3,4}$ can be obtained by applying the quantum channel $\mathscr{N}_3[\cdot]$ in Eq.~\eqref{eq:def_3_channel} to $H_{E,\gtri}$:
\begin{equation}
    H_{E,34} \sim \sum_{n=2}^{L-1} Z_n Z_{n+1} - \sum_{n=3}^{L} X_n,
\end{equation}
which describes an Ising chain on sites $2$ to $L$ with a mixed boundary at the left end. Because $H_{E,34}$ does not act on site $1$, there is an additional two-fold degeneracy, so the ES is at least four-fold degenerate. The ES obtained from the quantum channel and the wavefunction, shown in Figs.~\ref{Fig:ES_Ising_Z2_T}c and d, respectively, agree well with this analysis. The extra two-fold degeneracy is accidental and can be lifted by perturbing $H_{3,4}$ with 
$\sum_n \left( Z_n Y_{n+1} Z_{n+2} Y_{n+3} + Y_n Z_{n+1} Y_{n+2} Z_{n+3} \right)$~\cite{Ruben_SPT_and_transition_2017}.

\section{ES of $\mathbb{Z}_2\times \mathbb{Z}_2$ gSPT states with central charge $c=1$}\label{Sec:c=1_Z2_Z2}

We apply our framework  to gSPT states constructed from Luttinger liquid critical states with $c=1$ and continuously tunable criticality. 
The trivial $\mathbb{Z}_2\times\mathbb{Z}_2$ gSPT model with $c=1$  is~\cite{Scaffidi_gSPT_2017,Scaffidi_topo_luttinger_2021}:
\begin{equation}\label{eq:Z2_Z2_c_1_gtri}
    H^{\LL}_{\gtri}=-\sum_{n=1}^{N-2}\Big(X_{2n}X_{2n+2}+J Y_{2n}Y_{2n+2}+Z_{2n}Z_{2n+2}\Big) 
    - \sum_{n=1}^{N} X_{2n-1}.
\end{equation}
On the even sublattice, this is the spin-$\frac{1}{2}$ $XYX$ chain with anisotropy $J$. The $XYX$ and $XXZ$ chains are related by the on-site unitary transformation $U_X(\frac{\pi}{4})$. We consider $J \in (-1,1]$ to ensure the system is in a gapless Luttinger liquid (LL) phase. On the odd sublattice, the ground state is a product state.  The model has the $\mathbb{Z}_2\times \mathbb{Z}_2$ symmetry generated by $\prod_{n=1}^{N-1} X_{2n}$ and $\prod_{n=1}^{N} X_{2n-1}$, as well as a U(1) symmetry generated by $\sum_{n=1}^{N-1} Y_{2n}$. For simplicity, we fix the physical boundary spins of the $XYX$ chain by applying the projectors $(1+X_2)/2$ and $(1+X_{2N-2})/2$ to Eq.~\eqref{eq:Z2_Z2_c_1_gtri}. 

The ground state of $H^{\LL}_{\gtri}$ is 
\begin{equation}
    \ket{\Psi^{\LL}_\gtri} = \ket{+}_{\odd} \otimes \ket{\Psi_{\LL}}_{\even},
\end{equation}
where $\ket{\Psi_{\LL}}$ is the ground state of the critical $XYX$ chain with fixed boundary spins in the $X$ direction and ``LL'' denotes Luttinger liquid. Following the procedure outlined in Sec.~\ref{sec:c_half_z2_z2}, a family of non-trivial $\mathbb{Z}_2\times\mathbb{Z}_2$ gSPT states is obtained as
\begin{equation}\label{eq:psi_gSPT_c_1}
    \ket{\Psi^{\LL}_\gSPT(\theta)} = U_{\CZ} U_X(\theta) \ket{\Psi^{\LL}_\gtri}.
\end{equation}

Before discussing the ES of $\ket{\Psi^{\LL}_\gtri}$ and $\ket{\Psi^{\LL}_\gSPT(\theta)}$, let us briefly review the BCFT describing the $XYX$ chain. The critical phase of the $XYX$ chain is described by the compactified free boson CFT with central charge $c=1$~\cite{yellow_book_CFT}, whose action is 
\begin{equation}
    S = \frac{1}{8\pi} \int dx\, d\tau \left[ (\partial_x \phi)^2 + (\partial_\tau \phi)^2 \right],
\end{equation}
where the scalar field $\phi$ is compactified on a circle of radius $R$: $\phi \equiv \phi + 2\pi R$. The radius $R$ is related to the Luttinger parameter $K$ via $R = 2\sqrt{K}$, which is determined by the spin anisotropy as $K = \pi/[2\pi - 2\arccos(J)]$~\cite{quantum_1D_2003}. There is a dual field $\Theta$ defined by $\partial_\tau \phi = \partial_x \Theta$ and $\partial_x \phi = - \partial_\tau \Theta$, with dual radius $2/R$. In this free boson field theory, the boundary conditions are defined as follows: the Neumann (N) boundary is $\partial_x \phi(x,\tau)|_{x=0} = 0$, implying $\Theta(0,\tau) = \Theta_0$; while the Dirichlet (D) boundary is $\partial_x \Theta(x,\tau)|_{x=0} = 0$, implying $\phi(0,\tau) = \phi_0$.  

When both boundaries are Neumann (i.e., $\Theta$ is fixed), the BCFT spectrum is given by~\cite{Affleck_1992,Huang_2024}:
\begin{equation}\label{eq:boson_BCFT_spec_fix_fix}
    E_{\text{N,N}} \sim \frac{1}{2} \left( \frac{\Delta \Theta}{2\pi} + \frac{e R}{2} \right)^2 + n,
\end{equation}
where $\Delta \Theta$ is the difference between the $\Theta$ fields at the two boundaries, $e \in \mathbb{Z}$, and $n \in \mathbb{N}$. When one boundary is Neumann and the other is Dirichlet, the spectrum becomes
\begin{equation}\label{eq:boson_BCFT_spec_fix_free}
    E_{\text{D,N}} \sim n.
\end{equation}

From  bosonization one sees~\cite{Affleck_1992,quantum_1D_2003}, that applying a boundary field in the $XZ$ plane to the $XYX$ chain fixes the $\Theta$ field, resulting in Neumann boundary conditions; while applying a boundary field in the $Y$ direction fixes the $\phi$ field, resulting in Dirichlet boundary conditions. The usual open boundary of the $XYX$ chain corresponds to the Dirichlet boundary condition where $\phi(0,\tau)$ is fixed and $\Theta$ is free. This can be understood from the U(1) symmetry of the $XYX$ chain, which requires $\Theta \to \Theta + \text{constant}$ in the field theory. A boundary magnetic field in the $XZ$ plane explicitly breaks the U(1) symmetry, fixing $\Theta$ at the boundary. Conversely, open boundaries or a boundary field in the $Y$ direction preserve U(1) symmetry, leaving $\Theta$ free and $\phi$ fixed at the boundary.

We now analyze the ES of $\ket{\Psi^{\LL}_{\gtri}}$ and $\ket{\Psi^{\LL}_{\gSPT}(\theta)}$. 
For $\ket{\Psi^{\LL}_{\gtri}}$, since the physical boundary spins are fixed in the $X$ direction and the entanglement boundary does not break the U(1) symmetry, we obtain Neumann-Dirichlet boundary conditions and the bulk–boundary correspondence in Fig.~\ref{fig:bulk_boundary}a allows us to propose the following effective EH:
\begin{equation}\label{eq:EH_LL}
   H_{E,\gtri}^{\LL} \sim 
   P^{+}_{L}\!\left[\sum_{n=1}^{L-1}\left(
   X_{n}X_{n+1}+J Y_{n}Y_{n+1}+Z_{n}Z_{n+1}
   \right)\right]\!P^{+}_{L},
\end{equation}
where $P_L^{+}=(1+X_L)/2$.
According to Eq.~\eqref{eq:boson_BCFT_spec_fix_free}, the resulting ES consists of integer-spaced levels. The ES obtained from the iMPS simulations is fully consistent with this prediction, as shown in Fig.~\ref{Fig:ES_Z2_Z2_c_1}a. Technical details of the numerical simulations are provided in Appendix~\ref{app:OBC_IMPS}.

\begin{figure}
        \centering
        \includegraphics[width=\linewidth]{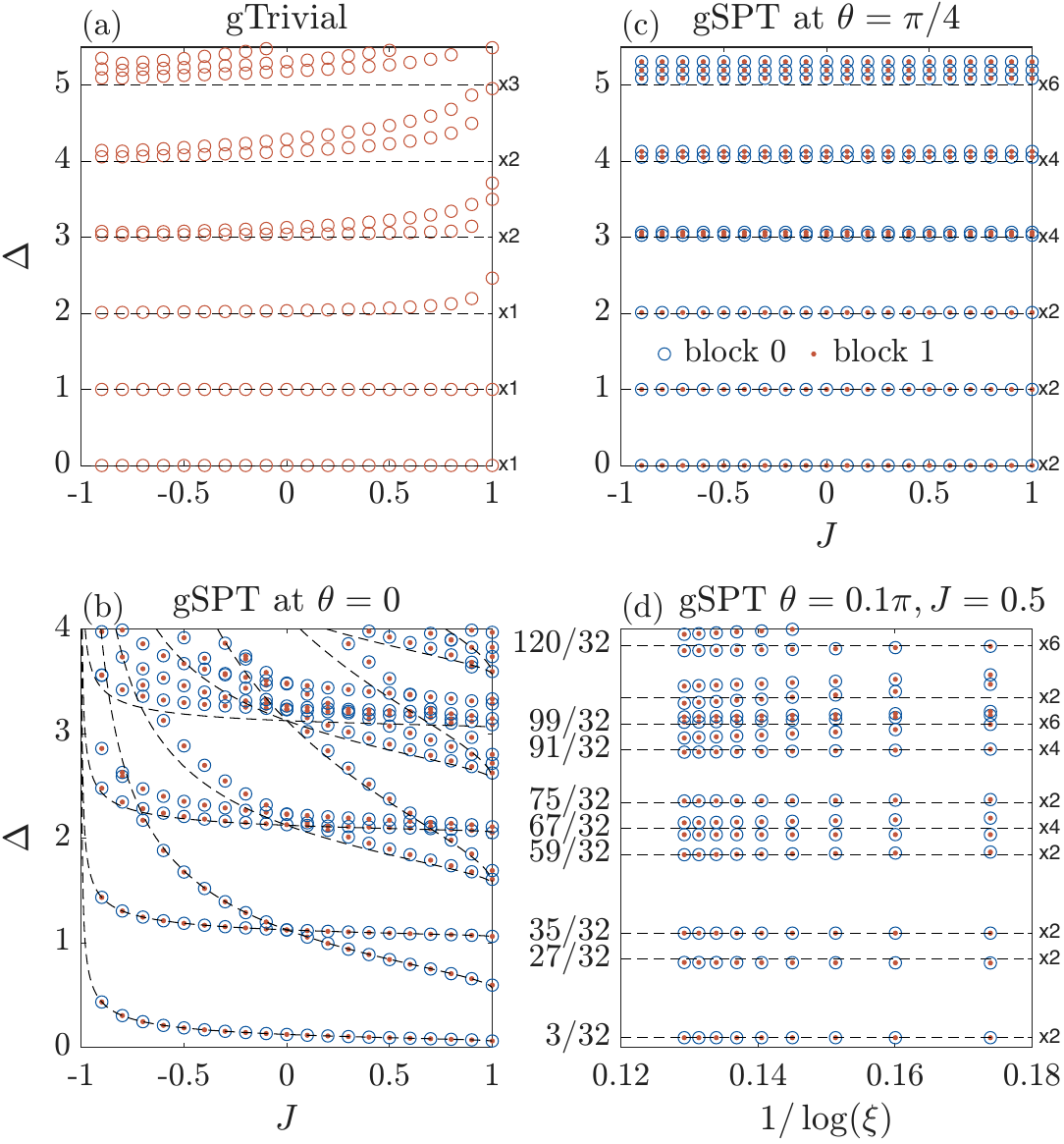}
   \caption{\textbf{Entanglement Spectra (ES) of the $\mathbb{Z}_2\times\mathbb{Z}_2$ gSPT states with $c=1$.} 
Panel (a) shows the ES extracted from iMPS with bond dimension $\chi=200$, approximating the ground states of the $XYX$ chain for varying $J$. 
The ES in panels (b) and (c) are obtained by applying the quantum channel to the iMPS in panel (a). 
Panel (d) shows the ES obtained by applying the quantum channel to iMPS of the $XYX$ chain at $J=0.5$, with bond dimensions ranging from $\chi=100$ to $500$. 
To directly compare with the CFT predictions (dashed lines) in Eqs.~\eqref{eq:boson_BCFT_spec_fix_fix} and~\eqref{eq:boson_BCFT_spec_fix_free}, the ES are shifted and rescaled. 
(a) ES of $\ket{\Psi^{\LL}_{\gtri}}$. 
(b) ES of $\ket{\Psi^{\LL}_{\gSPT}(0)}$. 
(c) ES of $\ket{\Psi^{\LL}_{\gSPT}(\frac{\pi}{4})}$. 
(d) ES of $\ket{\Psi^{\LL}_{\gSPT}(\frac{\pi}{10})}$ at $J=0.5$.
}
    \label{Fig:ES_Z2_Z2_c_1}
\end{figure}

Then we consider the ES of the non-trivial gSPT state $\ket{\Psi^{\LL}_{\gSPT}(\theta)}$. 
As in the previous cases, the reduced density matrices $\rho^{\LL}_{\gtri}$ of $\ket{\Psi^{\LL}_{\gtri}}$ and $\rho^{\LL}_{\gSPT}$ of $\ket{\Psi^{\LL}_{\gSPT}}$ are exactly related by a quantum channel:
\begin{align}\label{eq:channel4CZX_LL}
    \rho^{\LL}_{\gSPT}(\theta)\cong 
    \varrho^{\LL}_{\gSPT}(\theta)
    =\mathscr{N}_{CZ}\!\left[\mathscr{N}_X\!\left[\rho^{\LL}_{\gtri}\right]\right].
\end{align}
Analogous to the example in Sec.~\ref{sec:c_half_z2_z2}, when $\theta=0$ the spin nearest to the entanglement cut is fixed in the $Z$ direction. The corresponding effective EH becomes
\begin{equation}
   H_{E,\gSPT}^{\LL}\sim 
   P^{+}_{L}\!\left[
   \sum_{n=2}^{L-1}\left(X_{n}X_{n+1}+J Y_{n}Y_{n+1}\right)
   +\sum_{n=1}^{L-1}Z_{n}Z_{n+1}
   \right]\!P^{+}_{L}.
\end{equation}
As a result, both the entanglement boundary and the physical boundary correspond to Neumann boundary condition, implying that the dual field $\Theta$ is fixed at both boundaries. The difference between the boundary field values is $\Delta\Theta=\pi R/2$. Consequently, the low-lying ES is governed by Eq.~\eqref{eq:boson_BCFT_spec_fix_fix}. The numerical results shown in Fig.~\ref{Fig:ES_Z2_Z2_c_1}b are in excellent agreement with this prediction.

In contrast, when $\theta=\frac{\pi}{4}$ the spin nearest to the entanglement cut is fixed in the $Y$ direction. The corresponding effective EH is
\begin{equation}
   H_{E,\gSPT}^{\LL}\sim 
   P^{+}_{L}\!\left[
   \sum_{n=2}^{L-1}\left(X_{n}X_{n+1}+Z_{n}Z_{n+1}\right)
   +\sum_{n=1}^{L-1}J Y_{n}Y_{n+1}
   \right]\!P^{+}_{L}.
\end{equation}
In this case, the entanglement boundary corresponds to the Dirichlet boundary condition, while the physical boundary corresponds to the Neumann boundary condition. Consequently, the ES is still governed by Eq.~\eqref{eq:boson_BCFT_spec_fix_free}. However, the spectrum exhibits an additional twofold degeneracy arising from the non-trivial SPT order. The numerical results shown in Fig.~\ref{Fig:ES_Z2_Z2_c_1}c are consistent with this prediction.

What happens when $\theta\in(0,\frac{\pi}{4})$? 
We can again analyze this problem using boundary RG flow. It has been shown that, throughout the entire critical phase, the boundary entropy of the Dirichlet boundary is always higher than that of the Neumann boundary~\cite{Affleck_1998}.  
Since we have identified that at $\theta=0$ ($\theta=\frac{\pi}{4}$) the entanglement boundary condition is Neumann (Dirichlet), we expect that for $\theta\in(0,\frac{\pi}{4})$ the ES flows to the BCFT describing the ES at $\theta=0$, which has lower boundary entropy and is therefore more stable under the protecting symmetry. This scenario is the same as that for the $\mathbb{Z}_2\times\mathbb{Z}_2$ gSPT states with $c=\frac{1}{2}$ in Sec.~\ref{sec:c_half_z2_z2} and is illustrated Fig.~\ref{fig:RG}.
Our theoretical analysis is supported by the numerical ES of $\ket{\Psi_{\gSPT}(\frac{\pi}{10})}$ at $J=0.5$, shown in Fig.~\ref{Fig:ES_Z2_Z2_c_1}d.

\section{Non-invertible SPT and gSPT states}\label{sec:non_inv_SPT}

Our framework is based on MPUs and does not depend on the specific form of the protecting symmetry. It therefore applies to  non-invertible SPT states, provided they can be constructed by MPUs acting on a trivial state. A prominent example is given by the group-based cluster states~\cite{G_cluster_2015,Nat_2025}, which realize $\mathrm{Rep}(G)\times G$ SPT phases, where $G$ is a finite group. When $G$ is Abelian, $\mathrm{Rep}(G)\cong G$, and these states reduce to conventional cluster states. 
Denoting the local basis of a $G$ qudit by $\{\ket{g}\,|\, g\in G\}$, we define the operators~\cite{kitaev_2002}
\begin{align}
    \overrightarrow{X}^{[g]} \ket{h} &= \ket{gh}, \qquad
    \overleftarrow{X}^{[g]} \ket{h} = \ket{hg^{-1}}, \notag\\
    \overrightarrow{Z}^{[g]} \ket{h} &= \delta_{h,g}\ket{h}, \qquad
    \overleftarrow{Z}^{[g]} \ket{h} = \delta_{h^{-1},g}\ket{h}.
\end{align}
Here $\overrightarrow{X}^{[g]}$ and $\overleftarrow{X}^{[g]}$ are the right and left regular representations of $G$, generalizing the $\mathbb{Z}_2$ Pauli operator $X$. Using the irreducible representation matrices $\{\Gamma(g)\,|\, g\in G\}$, we define the transformed operators
$
Z^{[\Gamma]}_{\alpha\beta}
=
\sum_{g} \Gamma_{\alpha\beta}(g)\,\overrightarrow{Z}^{[g]}
=
\sum_{g} \Gamma_{\alpha\beta}(g)\ket{g}\bra{g},
$
which generalize the $\mathbb{Z}_2$ Pauli operator $Z$. And the Hadamard-like basis transformation for $G$ qudits is given by
$
\ket{\Gamma_{\alpha\beta}}
= \sqrt{\frac{d_{\Gamma}}{|G|}} \sum_{g} \Gamma_{\alpha\beta}(g)\ket{g}$,
where $d_{\Gamma}$ denotes the dimension of the irrep $\Gamma$.

The Hamiltonian of the trivial $\mathrm{Rep}(G)\times G$ SPT phase is given by
\begin{equation}\label{eq:tri_non_inv_SPT}
    H_{\text{tri}}
    = -\sum_{n}\sum_{\Gamma} d_{\Gamma}\,\mathrm{Tr}\, Z^{[\Gamma]}_{2n}
      -\sum_{n}\sum_{g}\overrightarrow{X}^{[g]}_{2n-1}.
\end{equation}
This model is invariant under a $\mathrm{Rep}(G)\times G$ symmetry generated by
\begin{equation}
    \mathrm{MPO}_{\Gamma}
    = \mathrm{Tr}\!\left(\prod_{n=1}^{M} Z^{[\Gamma]}_{2n}\right),
    \qquad
    U_{g}=\prod_{n} \overleftarrow{X}^{[g]}_{2n-1},
\end{equation}
where the MPO symmetry is non-invertible whenever $d_{\Gamma}>1$. 
The ground state of the trivial SPT Hamiltonian~\eqref{eq:tri_non_inv_SPT} is the product state
\begin{equation}\label{eq:prod_state}
    \ket{\Psi_{\text{tri}}}
= \ket{\pmb{1},e,\pmb{1},e,\ldots,\pmb{1},e},
\end{equation}
with $e\in G$ the identity element and $\pmb{1}\in \mathrm{Rep}(G)$ the trivial irrep of $G$.

A non-trivial $\mathrm{Rep}(G)\times G$ SPT state can be generated by acting on
$\ket{\Psi_{\text{tri}}}$ with a $\mathrm{Rep}(G)\times G$ symmetric MPU,
\begin{align}\label{eq:generalize_CX}
    U_{CX}
    = \prod_{n} C\overleftarrow{X}_{2n-2,\,2n-1}&
       \prod_{n} C\overrightarrow{X}_{2n-1,\,2n}, \notag\\
    C\overrightarrow{X}_{n,n+1}\ket{g_n,g_{n+1}}
    &= \overrightarrow{X}^{[g_n]}_{n+1}\ket{g_n,\, g_{n+1}}= \ket{g_n,\, g_n g_{n+1}}, \notag\\
    C\overleftarrow{X}_{n,n+1}\ket{g_n,g_{n+1}}
    &=\overleftarrow{X}^{[g_n]}_{n+1}\ket{g_n,\, g_{n+1}}= \ket{g_n,\, g_{n+1} g_n^{-1}},
\end{align}
where $C\overrightarrow{X}_{n,n+1}$ and $C\overleftarrow{X}_{n,n+1}$ denote the generalized
controlled-$X$ gates associated with the right and left regular representations of $G$,
respectively. The corresponding non-trivial $\mathrm{Rep}(G)\times G$ SPT Hamiltonian is
$H_{\text{SPT}} = U_{CX} H_{\text{tri}} U_{CX}^{\dagger}$,
and its ground state is given by $U_{CX}\ket{\Psi_{\text{tri}}}$.

We now demonstrate how to apply our framework to non-invertible SPT and gSPT states, using the example $G=S_3=\{e,r,\bar r,s,sr,s\bar r\}$, where the generators satisfy $s^2=1$, $r^3=1$, and $\bar r=r^2$. The irreducible representations of $S_3$ are $\Gamma\in\{\pmb{1},A,E\}$, with dimensions $d_{\pmb{1}}=d_A=1$ and $d_E=2$. Since the ES of gapped non-invertible SPT states have not been systematically studied, we first analyze the gapped case and then turn to non-invertible gSPT states.

Unlike the Abelian case, where all distinct gapped SPT phases can be generated by applying $U_{CX}^n$ with $n\in\mathbb{N}$ to a trivial SPT state, the non-Abelian case is more subtle: it remains unclear whether repeated applications of $U_{CX}$ generate all independent $\mathrm{Rep}(G)\times G$ SPT phases~\cite{Nat_2025}. Following the procedure outlined in previous sections, we consider an entanglement cut between sites $N-1$ and $N$. The reduced density matrix of the state $\ket{\Psi^{[p]}_{\mathrm{SPT}}}=U_{CX}^{p}\ket{\Psi_{\mathrm{Tri}}}$ is\footnote{Ignoring the generalized control-X gates that do not cross the entanglement cut, we have
\begin{align}
    \varrho_{G,\SPT}^{[p]}&=\Tr_{N-1}\left(\overrightarrow{CX}^p\ket{\pmb{1}e}_{N-1,N}\bra{\pmb{1}e}_{N-1,N}\overrightarrow{CX}^{\dagger p}\right)\notag\\
    &=\frac{1}{|G|^2}\Tr_{N-1}\sum_{g,h}\left[\left(\overrightarrow{X}_{N}^{[g]}\right)^p\ket{ge}_{N-1,N}\bra{he}_{N-1,N}\left(\overrightarrow{X}_N^{[h]\dagger}\right)^p\right]\notag\\
    &=\frac{1}{|G|}\sum_{g}\left[\left(\overrightarrow{X}_N^{[g]}\right)^p\ket{e}_{N}\bra{e}_{N}\left(\overrightarrow{X}_N^{[g]\dagger}\right)^p\right].\notag
\end{align}
}
\begin{equation}
      \rho^{[p]}_{G,\mathrm{SPT}}\cong\varrho^{[p]}_{G,\mathrm{SPT}}=\mathscr{N}^{[p]}_G[\rho_{G,\mathrm{tri}}]=\frac{1}{|G|}\sum_{g\in G}
    \left(\overrightarrow{X}^{[g]}_{N}\right)^{p}
    \rho_{G,\mathrm{tri}}
    \left(\overrightarrow{X}^{[g]}_{N}\right)^{\dagger p},
\end{equation}
where $\rho_{G,\mathrm{tri}}$ is the reduced density matrix of $\ket{\Psi_{\mathrm{tri}}}$. Taking the product state in Eq.~\eqref{eq:prod_state} into consideration, we have $\varrho^{[p]}_{G,\mathrm{SPT}}=\sum_{g\in G}\lambda^{[p]}_g\ket{g}\bra{g}$, where $\lambda^{[p]}_g = \left| \{ h \mid h \in G, h^p = g \} \right|/|G|$ with the numerator being the multiplicity of $g\in G$ in the multiset $G^{[p]}=\{h^p \mid h\in G\}$.
For $G=S_3$, the vectors $\lambda^{[p]}=(\lambda^{[p]}_g,\lambda^{[p]}_h,\cdots)$ are
\begin{align}
    &\lambda^{[1]}=\left(\tfrac{1}{6},\tfrac{1}{6},\tfrac{1}{6},\tfrac{1}{6},\tfrac{1}{6},\tfrac{1}{6}\right), \qquad
    \lambda^{[2]}=\left(\tfrac{2}{3},\tfrac{1}{6},\tfrac{1}{6},0,0,0\right), \notag\\
    &\lambda^{[3]}=\left(\tfrac{1}{2},0,0,\tfrac{1}{6},\tfrac{1}{6},\tfrac{1}{6}\right), \qquad
    \lambda^{[4]}=\left(\tfrac{2}{3},\tfrac{1}{6},\tfrac{1}{6},0,0,0\right), \notag\\
    &\lambda^{[5]}=\left(\tfrac{1}{6},\tfrac{1}{6},\tfrac{1}{6},\tfrac{1}{6},\tfrac{1}{6},\tfrac{1}{6}\right), \qquad
    \lambda^{[6]}=\left(1,0,0,0,0,0\right).
\end{align}
In particular, it implies the ES obtained from the quantum channel at $p=1$ exhibits a sixfold degeneracy, in agreement with the numerical results reported in Ref.~\cite{Nat_2025}. 

\begin{figure}
    \centering
    \includegraphics[width=\linewidth]{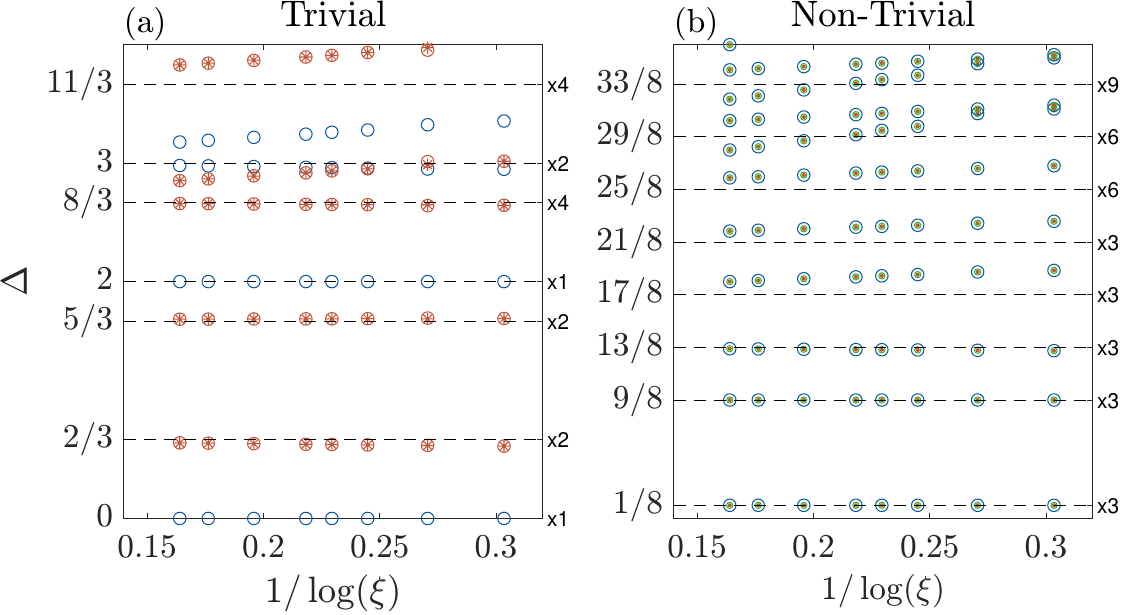}
    \caption{\textbf{Entanglement spectra (ES) of the $\mathbb{Z}_3\times\mathbb{Z}_3$ gSPT states with $c=4/5$.}
All ES are extracted from $\mathbb{Z}_3$-symmetric iMPS~\cite{Xu_2017} (corresponding to free physical boundaries), with bond dimensions $\chi$ in the range $30\le\chi\le200$. 
(a) ES of the trivial $\mathbb{Z}_3\times\mathbb{Z}_3$ gSPT state, which is identical to that of the trivial Rep$(S_3)\times S_3$ gSPT state; different markers denote the $\mathbb{Z}_3$ charges of the iMPS. 
(b) ES of the non-trivial $\mathbb{Z}_3\times\mathbb{Z}_3$ gSPT state, where different markers correspond to eigenvalues of the $\mathbb{Z}_3$ operator  $\hat{X}_N$, see Eq.~\eqref{eq:S_3_channel}. 
Multiplying the level degeneracies by a factor of two yields the ES of the non-trivial Rep$(S_3)\times S_3$ gSPT state.}
    \label{Fig:Z3_ES_conserve}
\end{figure}

Next, let us consider the ES of the Rep$(S_3)\times S_3$ gSPT states, and we only focus on the MPU $U_{CX}$, i.e., $p=1$. Similar to the $\mathbb{Z}_2\times\mathbb{Z}_2$ cluster chain, the $\mathrm{Rep}(S_3)\times S_3$ SPT models can be tuned to phase transition points by adding a $\mathrm{Rep}(G)$-symmetric term
$-J\sum_n\sum_{g}\overleftarrow{X}^{[g]}_{2n}\overrightarrow{X}^{[g]}_{2n+2}
$
on the even sublattice, respectively. 
For sufficiently large $J$, the $\mathrm{Rep}(G)$ symmetry is spontaneously broken. 
However, for $G=S_3$, the phase transition from the $\mathrm{Rep}(S_3)$-symmetric phase to the spontaneously symmetry-broken phase is first order~\cite{Sergej_2025}, which excludes the existence of a $\mathrm{Rep}(S_3)\times S_3$ gSPT state along this path. To obtain a continuous phase transition, we instead add a different term on the even sublattice to $H_{\text{tri}}$ and $H_{\SPT}$ where we consider only the $r$ and $\bar r$ elements:
\begin{align}\label{eq:H_rep_S3_2_Z2}
H_{\gtri}
&=H_{\text{tri}}
-J\sum_n\bigl(
\overleftarrow{X}^{[r]}_{2n}\overrightarrow{X}^{[r]}_{2n+2}
+\overleftarrow{X}^{[\bar r]}_{2n}\overrightarrow{X}^{[\bar r]}_{2n+2}
\bigr),\notag\\
H_{\gSPT}
&=U_{CX}H_{\gtri}U_{CX}^{\dagger}.
\end{align}
For sufficiently large $J$, this new term spontaneously breaks the $\mathrm{Rep}(S_3)$ symmetry down to $\mathrm{Rep}(\mathbb{Z}_2)\cong \mathbb{Z}_2$~\cite{Xu_2022}. 
The $6\times6$ matrices $\overleftarrow{X}^{[g]}$, $\overrightarrow{X}^{[g]}$, and $Z^{[\Gamma]}_{\alpha\beta}$ can be expressed in terms of $\mathbb{Z}_2$ Pauli operators $\{Z,X\}$ and $\mathbb{Z}_3$ clock operators $\{\hat{Z},\hat{X}\}$,\footnote{The operators $\overrightarrow{X}^{[g]}$, $\overleftarrow{X}^{[g]}$, and $Z^{[\Gamma]}$ depend on the choice of the group $G$. We do not explicitly show $G$-dependence and take $G=S_3$ in the examples. For $G=\mathbb{Z}_3=\{1,a,a^2\}$, the left and right regular representations coincide, $\overrightarrow{X}^{[a]}=\overleftarrow{X}^{[a]}$, and we denote them by $\hat{X}$. Denoting the irreducible representations of $\mathbb{Z}_3$ by $\{[1],[\omega],[\bar{\omega}]\}$, where $\omega=\exp(i2\pi/3)$, we define $Z^{[\omega]} \equiv \hat{Z}$.} under which $H_{\gtri}$ takes the form
\begin{align}
&H_{\gtri}
=-\sum_{n\in\even}\frac{1+Z_n}{2}\frac{1+\hat Z_n+\hat Z_n^{\dagger}}{3}
-\sum_{n\in\odd}\frac{1+X_n}{2}\frac{1+\hat X_n+\hat X_n^{\dagger}}{3}\notag\\
&-J\sum_{n\in\even}\Bigl[
\frac{1+Z_n}{2}\bigl(\hat X_n^{\dagger}\hat X_{n+2}+\mathrm{h.c.}\bigr)
+\frac{1-Z_n}{2}\bigl(\hat X_n\hat X_{n+2}+\mathrm{h.c.}\bigr)
\Bigr].
\end{align}
In the ground-state subspace, $Z_{2n}=1$ and $X_{2n+1}=1$. 
Consequently, the Hamiltonian on the even sublattice reduces to the $\mathbb{Z}_3$ clock model, while the Hamiltonian on the odd sublattice is trivial. 
The phase transition of the models in Eq.~\eqref{eq:H_rep_S3_2_Z2} therefore occurs at $J=1$ and is described by a CFT with central charge $c=4/5$~\cite{yellow_book_CFT}. 
Then the ground state ES is governed by the associated BCFT~\cite{Potts_BCFT_1998}.

Using the generalized Hadamard transformation, the generalized projectors can be obtained from the regular representations
\begin{equation}
    P^{[\Gamma]}_{\alpha\beta,N}
=\sqrt{\frac{d_{\Gamma}}{|G|}}\sum_{g\in G}\Gamma_{\alpha\beta}(g)\overrightarrow{X}^{[g]}=\sqrt{\frac{1}{d_{\Gamma}}}\sum_{\eta}
\ket{\Gamma_{\alpha\eta}}_N\bra{\Gamma_{\beta\eta}}_N,
\end{equation}
which are idempotent for $\alpha=\beta$ and nilpotent for $\alpha\neq\beta$. 
Under this transformation, the quantum channel relation ($p=1$) between reduced density matrices becomes
\begin{equation}
\varrho_{G,\gSPT}
=\sum_{\alpha\beta\Gamma} P^{[\Gamma]}_{\alpha\beta,N}
\rho_{G,\gtri} P^{[\Gamma]\dagger}_{\alpha\beta,N}.
\end{equation}
Expressing both $\rho_{S_3,\gtri}$ and $P^{[\Gamma]}_{\alpha\beta,N}$ in terms of $\mathbb{Z}_2$ and $\mathbb{Z}_3$ operators, we have
\begin{equation}\label{eq:S_3_channel}
\varrho_{S_3,\gSPT}
=\frac{\mathbbm{1}_2}{2}\otimes
\left(
\sum_{p\in\{1,\omega,\bar{\omega}\}}P^{[p]}_N\rho_{\mathbb{Z}_3,\gtri} P^{[p]}_N
\right),\quad P^{[p]}=\ket{\hat{p}_X}\bra{\hat{p}_X},
\end{equation}
where $\rho_{\mathbb{Z}_3,\gtri}$ is the reduced density of the trivial $\mathbb{Z}_3\times\mathbb{Z}_3$ gSPT state, $\omega=e^{2i\pi/3}$
and $\ket{\hat{p}_X}$ are eigenstates of the $\mathbb{Z}_3$ operator $\hat{X}$.
Therefore, the BCFT describing the ES of the $\mathrm{Rep}(S_3)\times S_3$ gSPT state is almost identical to that of the $\mathbb{Z}_3\times\mathbb{Z}_3$ gSPT state. The only difference is an additional two-fold degeneracy originating from the identity operator $\mathbbm{1}_2$.

Fig.~\ref{Fig:Z3_ES_conserve}a shows the ES of the trivial $\mathbb{Z}_3\times\mathbb{Z}_3$ gSPT state with free physical boundaries, which agrees with the BCFT prediction~\cite{Affleck_1992,Huang_2025}. 
The trivial $\mathrm{Rep}(S_3)\times S_3$ gSPT state exhibits the same ES.
Fig.~\ref{Fig:Z3_ES_conserve}b shows the ES of the non-trivial $\mathbb{Z}_3\times\mathbb{Z}_3$ gSPT state with free physical boundaries. 
The ES in Figs.~\ref{Fig:Z3_ES_conserve}a and b are described by different BCFTs, since the MPU modifies the entanglement boundary condition. 
Multiplying the degeneracy of the ES in Fig.~\ref{Fig:Z3_ES_conserve}b by a factor of two yields the ES of the non-trivial $\mathrm{Rep}(S_3)\times S_3$ gSPT state.

\section{Conclusion and Outlook}\label{sec:conclusion_discusion}

In this work, we have developed a systematic framework for analyzing and understanding entanglement spectra of gapless Symmetry Protected Topological (gSPT) states in one spatial dimension. We show that the reduced density matrices of non-trivial gSPT states can be obtained exactly or approximately by applying quantum channels to the reduced density matrices of trivial gSPT states. The Kraus operators of these quantum channels can usually be expressed as projectors, which allows us to identify the corresponding conformal boundary conditions at the entanglement boundary. Together with the bulk–boundary correspondence, this provides a systematic way to deduce effective entanglement Hamiltonians and entanglement spectra. We furthermore find that, for a given gSPT phase, the entanglement boundary condition is not uniquely determined, and it can be different at some fine-tuned points. We identify the most stable one using the RG flow of boundary entropies.

Since we focus on gSPT states that are constructed using SPT entanglers, the gSPT states considered here are non-intrinsic gSPT states. 
The intrinisic gSPT states can be constructed using non-invertible transformations, such as the Kennedy--Tasaki transformation~\cite{KT_1992,KT_4_gSPT_2025}. In particular, some gSPT states whose protecting symmetry is also non-invertible can be constructed via the Kennedy--Tasaki transformation~\cite{Shao_2024,Huang_2025}. It would be interesting to investigate how non-invertible transformations, including the Kramers--Wannier transformation, modify the entanglement.
Moreover, since gapped and gapless SPT states, as well as symmetry-enriched topological states in higher dimensions~\cite{Lukas_2023}, can be generated using SPT entanglers, our framework can be naturally generalized to higher dimensions, and their effective reduced density matrices can be similarly obtained from those of trivial states by applying matrix-product-state quantum channels~\cite{MPQC_2026}. 
Furthermore, our findings offer a practical pathway for the experimental study of gSPT states starting from trivial gapless states, which have been recently adiabatically prepared in finite size systems \cite{experimental_CFT_spec_2026}. 
Since the required quantum channels are physically realizable—and in many scenarios, the Kraus operators simplify to projectors—it is possible to directly implement the channel via projective measurements, which can be experimentally advantageous compared to evolving the state with an SPT entangler.

\textbf{Acknowledgments.} We thank Rui-Zhen Huang and Ruben Verresen for insightful discussions. Calculations were performed using the TeNPy Library~\cite{tenpy}. We acknowledge support from the Deutsche Forschungsgemeinschaft (DFG, German Research Foundation) under Germany’s Excellence Strategy–EXC–2111–390814868, TRR 360 – 492547816 and DFG grants No. KN1254/1-2, KN1254/2-1, FOR 5522
(project-id 499180199), the European Union (grant agreement No 101169765), as well as the Munich Quantum Valley, which is supported by the Bavarian state government with funds from the Hightech Agenda Bayern Plus.

\textbf{Data availability.}
Data and codes are available upon reasonable request on Zenodo~\cite{zenodo}.

\appendix

\addtocontents{toc}{\string\tocdepth@munge}

\section{Additional details on the framework}\label{app:channel}
In this section, we review key properties of MPUs that lead to the quantum channel relation between density matrices, Eq.~\eqref{eq:channel}. We also show that $\rho_{\gSPT}$ and $\varrho_{\gSPT}$ exhibit the same projective symmetry. 

The reduced density matrix of $\ket{\Psi_{\gSPT}} = U \ket{\Psi_{\gtri}}$ can be expressed as
\begin{equation}\label{eq:MPU_to_QC}
\vcenter{\hbox{\includegraphics[width=\linewidth]{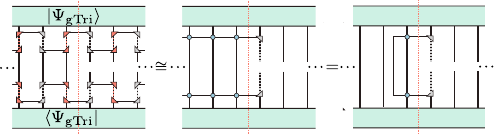}}}.
\end{equation}
In this graphical representation one needs to take the complex conjugate and transpose the physical indices of the MPU tensors in the bottom part. In the step marked by ``$\cong$'', we discard the gates acting on the right part of the system and transform the gates on the left part back into MPU tensors. For the second equality, we use the following pulling-through relation of MPU tensors~\cite{MPU_cirac_2017,MPU_Chen_2018}:
\begin{equation}
\vcenter{\hbox{\includegraphics[width=2cm]{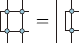}}},
\end{equation}
where we adopted a gauge in which the left fixed point of the MPU is the identity matrix,
\begin{equation}
\label{Eq:Eq_MPU_fixedpoint}
\vcenter{\hbox{\includegraphics[width=1.5cm]{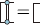}}}.
\end{equation}
On the right-hand side of Eq.~\eqref{eq:MPU_to_QC}, we cannot pull through the MPU tensors anymore due to the entanglement cut. We now consider the following decomposition~\cite{MPU_Chen_2018,MPU_cirac_2017}:
\begin{equation}\label{Eq:Eq_MPU_decomp}
\vcenter{\hbox{\includegraphics[width=2cm]{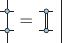}}} \quad + \text{other terms}.
\end{equation}
Substituting Eq.~\eqref{Eq:Eq_MPU_decomp} into Eq.~\eqref{eq:MPU_to_QC} and using Eq.~\eqref{Eq:Eq_MPU_fixedpoint}, we obtain
\begin{equation}\label{eq:rho_equal_varrho_sigma}
\vcenter{\hbox{\includegraphics[width=6cm]{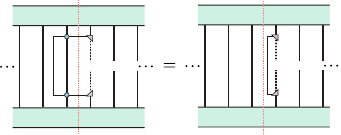}}} \quad + \varsigma,
\end{equation}
where the first term on the right-hand side is $\varrho_{\gSPT}$ and $\varsigma$ is the contribution obtained from the other terms in Eq.~\eqref{Eq:Eq_MPU_decomp}. Using the left-gauge condition of the MPU, we have
\begin{equation}
\vcenter{\hbox{\includegraphics[width=1cm]{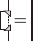}}}\; ,
\end{equation}
which implies that $\tr(\rho_{\gSPT}) = \tr(\varrho_{\gSPT})$. Consequently, $\tr(\varsigma) = 0$.

A more explicit expression for $\varsigma$ can be obtained by considering a class of SPT entanglers constructed from 2-cocycles of an Abelian group $G$~\cite{MPU_gong_2020}:
\begin{equation}
U=\sum_{\{g_1,\cdots,g_M\}}
\prod_{n=1}^M
\omega(g_n^{-1}h_{n+1},h_{n+1}^{-1})
\ket{g_1,\cdots, g_M}\bra{g_1,\cdots,g_M},
\end{equation}
where $g \in G$ and $\omega(g,h)$ is a 2-cocycle of $G$. The graphical representation of the MPU is
\begin{equation}\label{eq:MPU_SPT}
\vcenter{\hbox{\includegraphics[width=5cm]{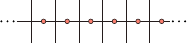}}},
\end{equation}
where the tensors at the crossing points are $\delta$ tensors, and the matrix elements represented by the red dots are $\omega(g_n^{-1}h_{n+1},h_{n+1}^{-1})$. For such MPU tensors, the decomposition~\eqref{Eq:Eq_MPU_decomp} takes the form
\begin{equation}\label{eq:decomp}
\vcenter{\hbox{\includegraphics[width=5cm]{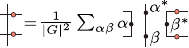}}},
\end{equation}
where $\alpha$ and $\beta$ are diagonal matrices, with $\alpha=\sum_{g\in G}\alpha(g)\ket{g}\bra{g}$ and $\alpha(g)$ denoting irreducible representations of $G$.
Substituting this decomposition into Eq.~\eqref{eq:MPU_to_QC}, we obtain
\begin{equation}
\rho_{\gSPT}\cong
\vcenter{\hbox{\includegraphics[width=7.5cm]{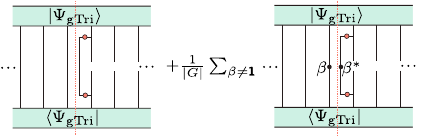}}},
\end{equation}
where the first term corresponds to $\varrho_{\gSPT}$ and the second term is $\varsigma$. In the $\varsigma$ term, $\beta$ acts on the traced physical degrees of freedom adjacent to the entanglement cut. If these degrees of freedom are unentangled from the rest of the system and transform according to the trivial irreducible representation of $G$, $\varsigma$ will vanish exactly.

We now analyze the symmetry properties of the reduced density matrices. 
They satisfy the following symmetry relations~\cite{MPU_gong_2020}:
\begin{equation}\label{eq:Eq_MPU_proj_sym}
    \vcenter{\hbox{\includegraphics[width=5cm]{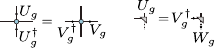}}},
\end{equation}
where $U_g$ is a linear representation acting on the physical degrees of freedom, while $V_g$ and $W_g$ are projective representations. We next show that both $\rho_{\gSPT}$ [the left-hand side of Eq.~\eqref{eq:rho_equal_varrho_sigma}] and $\varrho_{\gSPT}$ [the first term on the right-hand side of Eq.~\eqref{eq:rho_equal_varrho_sigma}] transform in the same manner under the symmetry $G$. 
To this end, we act with $W_g \otimes U_g \otimes U_g \otimes \cdots$ on the reduced density matrix $\rho_{\gSPT}$ (up to local unitary gates) and use the relations in Eq.~\eqref{eq:Eq_MPU_proj_sym} to track the transformation. This yields
\begin{equation}
    \vcenter{\hbox{\includegraphics[width=6.5cm]{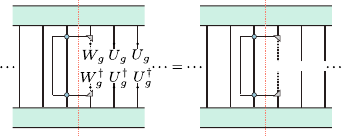}}},
\end{equation}
which shows that the symmetry $G$ acts projectively on $\rho_{\gSPT}$. Similarly, the action of $G$ on $\varrho_{\gSPT}$ is
\begin{equation}
    \vcenter{\hbox{\includegraphics[width=6.5cm]{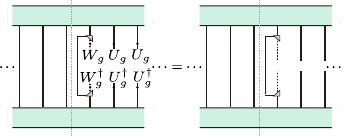}}},
\end{equation}
which is identical to that on $\rho_{\gSPT}$. We therefore conclude that neglecting the $\varsigma$ term does not change the symmetry properties of the reduced density matrices. Since the symmetry acts projectively on both $\rho_{\gSPT}$ and $\varrho_{\gSPT}$, the degeneracy of the ES follows directly.

\section{SPT entanglers as translational operators}\label{app:MPU_of_translation}

In this section, we show that the framework also applies when the MPU index is non-zero. The MPU index is defined through the following decomposition of the MPU tensor:
\begin{equation}\label{eq:MPU_SVD}
    \vcenter{\hbox{\includegraphics[width=4 cm]{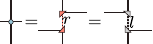}}},
\end{equation}
where $r$ and $l$ are the bond dimensions obtained from the decomposition. The index is given by $\frac{1}{2}\ln\frac{r}{l}$~\cite{Gross2012,MPU_Chen_2018,MPU_cirac_2017} and characterizes the amount of information flowing across a cut of the MPU. It is known that all finite-depth quantum circuits have zero index, whereas translational operators can have non-zero indices~\cite{MPU_Chen_2018}.  

There exist gSPT states related by translational symmetry~\cite{Gapless_SPT_Ruben_2021,Ruben_2025}. For example, the following $\mathbb{Z}_2\times\mathbb{Z}_2$ gSPT models with central charge $c=1/2$, which differ from those discussed in Sec.~\ref{sec:c_half_z2_z2}, are:
\begin{align}\label{eq:gSPT_translation}
    \tilde{H}_{\gtri}&=\sum_{n} X_n X_{n+1} + \sum_{n} (X_{2n-1} X_{2n} + Y_{2n-1} Y_{2n}), \notag\\
    \tilde{H}_{\gSPT}&=\sum_{n} X_n X_{n+1} + \sum_{n} (X_{2n} X_{2n+1} + Y_{2n} Y_{2n+1}).
\end{align}
The $\mathbb{Z}_2\times\mathbb{Z}_2$ symmetry of these models is generated by $\prod_n X_n$ and $\prod_n Y_n$. Here, one should treat the qubits at positions $2n-1$ and $2n$ as belonging to the same site to clearly distinguish the trivial and non-trivial gSPT phases.

Here, $H_{\gtri}$ and $H_{\gSPT}$ in Eq.~\eqref{eq:gSPT_translation} are related by translational operator, which can be represented as an MPU:
\begin{equation}
     \vcenter{\hbox{\includegraphics[width=6.5 cm]{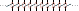}}},
\end{equation}
where the red ovals denote the MPU tensors and the lines represent identity matrices. We can apply the decomposition in Eq.~\eqref{eq:MPU_SVD} to the tensors of the translational MPU. The second equality in Eq.~\eqref{eq:MPU_SVD} is straightforward, giving $r=1$. The first equality can be written as
\begin{equation}
     \vcenter{\hbox{\includegraphics[width=2 cm]{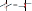}}} 
     = \frac{1}{2}\left(\mathbbm{1}_2 \otimes \mathbbm{1}_2 + X \otimes X - Y \otimes Y + Z \otimes Z\right),
\end{equation}
so that $l=4$, and the MPU index is $-\ln 2$. The standard form of the MPU is then
\begin{equation}
     \vcenter{\hbox{\includegraphics[width=6.5 cm]{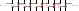}}}.
\end{equation}
It can be seen that, in this case, for any states related by the translational MPU, their reduced density matrices are exactly related via the quantum channel:
\begin{equation}
     \vcenter{\hbox{\includegraphics[width=8 cm]{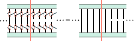}}}.
\end{equation}
This quantum channel also traces out the qubit at the right side of the entanglement cut. 

From the bulk-boundary correspondence, the EH of $\tilde{H}_{\gtri}$ in Eq.~\eqref{eq:gSPT_translation} is
\begin{equation}
    \tilde{H}_{E,\gtri} = \sum_{n=2}^{2L-2} X_n X_{n+1} + \sum_{n=1}^{L} (X_{2n-1} X_{2n} + Y_{2n-1} Y_{2n}).
\end{equation}
Using the fact that $[X_{2n-1}X_{2n},\tilde{H}_{E,\gtri}] = 0$, one can see that $\tilde{H}_{E,\gtri}$ is equivalent to an Ising chain with the free boundary condition at both ends.  

The quantum channel relation then implies that the EH of $\tilde{H}_{\gSPT}$ is obtained by tracing out site $1$:
\begin{equation}
    \tilde{H}_{E,\gSPT} = \sum_{n=2}^{2L-2} X_n X_{n+1} + \sum_{n=2}^{L} (X_{2n-1} X_{2n} + Y_{2n-1} Y_{2n}).
\end{equation}
This EH is equivalent to an Ising chain with a mixed left boundary and a free right boundary, implying the entanglement boundary condition is mixed while the physical boundary condition is free. We have numerically verified that this analysis is correct, indicating that our framework remains valid also when the MPU has a non-trivial index.

\section{Details on iMPS simulations }\label{app:OBC_IMPS}

To compute the ES, one can calculate the ground state $\ket{\Psi_\gtri}$ of the trivial gSPT model with OBC using the density matrix renormalization group (DMRG) method with site-dependent tensors~\cite{DMRG_1992}. Because the length of the BCFT describing the ES  scales logarithmically with the subsystem size~\cite{Calabrese_Cardy_2004,Calabrese_cardy_2009}, one has to simulate very long OBC chains. However, the variational uniform matrix product states (VUMPS) approach~\cite{VUMPS_2018,VUMPS_2019} is often more convenient, as it directly simulates the system in the thermodynamic limit $N \rightarrow +\infty$. This allows one to exploit translational symmetry and optimize only the tensors within a single unit cell.

Spin chains with different physical boundary conditions can be simulated by imposing appropriate symmetries on iMPS~\cite{Huang_2024}. We perform VUMPS simulations using the TenPy library, both with and without imposing symmetries on the iMPS. From the symmetry perspective, this approach is straightforward, while a more precise understanding from CFT can be found in Refs.~\cite{Cho_ES_BCFT_PRB_2017,Huang_2024}.  As an example, consider the Ising chain. The free OBC chain is 
\begin{equation}
    H_{\text{free}} = -\sum_{n=1}^{M-1} Z_n Z_{n+1} - \sum_{n=1}^{M} X_n,
\end{equation}
which has a global $\mathbb{Z}_2$ symmetry $\prod_{n=1}^M X_n$ and a unique ground state. To simulate the free OBC chain, we impose this global symmetry on the iMPS; otherwise, the iMPS may spontaneously break the symmetry at the critical point due to finite bond dimension truncation.  The mixed OBC chain is 
\begin{equation}
    H_{\text{mixed}} = -\sum_{n=1}^{M-1} Z_n Z_{n+1} - \sum_{n=2}^{M-1} X_n.
\end{equation}
It not only preserves the global $\mathbb{Z}_2$ symmetry but also has edge modes: $[H_{\text{mixed}}, Z_1]=[H_{\text{mixed}}, Z_M]=0$, which spontaneously break the global symmetry at the critical point. As a result, the mixed OBC chain has two exactly degenerate ground states. The fixed OBC chain is obtained by adding a boundary field,
\begin{equation}
    H_{\text{fixed}} = H_{\text{mixed}} - h_{\text{bdry}}(Z_1 + Z_M),
\end{equation}
which selects one of the two degenerate ground states. To simulate the ground state ES of a fixed OBC chain, we use iMPS without imposing the global symmetry, allowing the finite bond dimension truncation to induce the spontaneous $\mathbb{Z}_2$ symmetry breaking.

Another example is the $XYX$ chain. For the $XYX$ chain with the usual OBC,
\begin{equation}
    H^{\LL}_{\free}=\sum_{n=1}^{M-1}\left(X_nX_{n+1}+J Y_nY_{n+1}+Z_nZ_{n+1}\right),
\end{equation}
it preserves the U(1) symmetry and the ground state is unique. Therefore, the ES of this ground state can be simulated by imposing the U(1) symmetry on the iMPS. For the $XYX$ chain with the boundary spins fixed in the $X$ direction, the Hamiltonian as well as the ground state breaks the U(1) symmetry. The ES of the such fixed OBC ground state can thus be simulated using iMPS without imposing any symmetry constraint. The reason is that although the critical phase of the $XYX$ chain does not exhibit true long-range order, finite bond-dimension truncation induces spontaneous U(1) symmetry breaking, causing the spins to point along a particular direction in the $XZ$ plane. This orientation can be fixed to the $X$ direction by using real iMPS.

Other OBC spin chains can be simulated using iMPS in the same manner. The physics captured by iMPS and finite MPS is equivalent; the only difference is that finite-size scaling in finite MPS is replaced by finite-entanglement scaling or finite correlation length scaling in iMPS~\cite{Pollmann_finite_ent_scaling_2009}. Specifically, the system size $N$ is replaced by the iMPS correlation length $\xi$.

Since a non-trivial gSPT states is obtained by applying an MPU to a trivial gSPT state, the ES of a non-trivial gSPT state can be directly extracted from the iMPS of the trivial gSPT state. The mixed-gauge iMPS can be written as
\begin{equation}
\vcenter{\hbox{\includegraphics[width=5 cm]{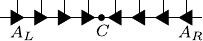}}},
\end{equation}
where $A_L$ and $A_R$ are isometric MPS tensors and $C$ is the central tensor. The effective reduced density matrix of the iMPS relevant for computing the ES is given by $C C^{\dagger}$. Suppose that the iMPS approximates the ground state of the trivial gSPT model. When $\varsigma = 0$, according to Eq.~\eqref{eq:rho_equal_varrho_sigma}, the ES of the non-trivial gSPT state can be obtained from
\begin{equation}
\varrho_{\gSPT}
\cong
\vcenter{\hbox{\includegraphics[width=1 cm]{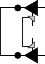}}},
\end{equation}
where the rows (columns) of the matrix correspond to the combined open physical and virtual indices of the bra (ket), respectively. When $\varsigma \neq 0$, the exact ES of the non-trivial gSPT state is obtained from
\begin{equation}
\rho_{\gSPT}
\cong
\vcenter{\hbox{\includegraphics[width=1.4 cm]{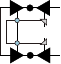}}}.
\end{equation}

\section{Smooth $\mathbb{Z}_2\times\mathbb{Z}_2^T$ symmetric path connecting gSPT states with and without gapped sectors}\label{app:same_phase}

Based on the definition that distinct gSPT phases cannot be connected by symmetry-preserving paths without phase transitions, we show in this section that certain $\mathbb{Z}_2 \times \mathbb{Z}_2^T$ gSPT states with and without gapped sectors nevertheless belong to the same gSPT phase.

Consider a $\mathbb{Z}_2 \times \mathbb{Z}_2^T$-symmetric state
$\ket{\Psi_1} = \prod_n CZ_{n,n+1}\ket{\Psi_0}$,
where $\ket{\Psi_0}$ is a trivial $\mathbb{Z}_2 \times \mathbb{Z}_2^T$ gSPT state in the thermodynamic limit and without a gapped sector. 
Denoting the reduced density matrix of $\ket{\Psi_0}$ by $\rho_0$, the reduced density matrix of $\ket{\Psi_1}$ satisfies
$
\rho_1 \cong \mathscr{N}_{CZ}[\rho_0] + \varsigma_1
= \varrho_1 + \varsigma_1 .
$
Applying Stinespring’s dilation (Fig.~\ref{Fig:Stinespring_Dilation}) to the relation $\varrho_1 = \mathscr{N}_{CZ}[\rho_0]$, we obtain a state
$
\ket{\Phi_1}
= \prod_n CZ_{n,n+1}
\left(\ket{+}_{\odd}\otimes\ket{\Psi_0}_{\even}\right),
$
which contains a gapped sector and whose reduced density matrix is exactly equivalent to $\varrho_1$. 
We now show that $\ket{\Phi_1}$ and $\ket{\tilde{\Psi}_1}=\ket{+}_{\odd}\otimes\ket{\Psi_1}_{\even}$ can be connected by a $\mathbb{Z}_2 \times \mathbb{Z}_2^T$-symmetric circuit.

The two states $\ket{\tilde{\Psi}_1}$ and $\ket{\Phi_1}$ are related by the circuit
$
\ket{\tilde{\Psi}_1}
=
\prod_m CZ_{2m,2m+2}
\prod_n CZ_{n,n+1}
\ket{\Phi_1}.
$
Since all $CZ$ gates commute, the circuit can be rearranged into a product of three-site gates
\begin{align}
\mathcal{U}_{2n,2n+2}
&=
CZ_{2n,2n+1}
CZ_{2n+1,2n+2}
CZ_{2n,2n+2}
\notag\\
&=
\frac{1}{8}
\bigl(
Z_{2n}
+
Z_{2n+1}
+
Z_{2n+2}
-
Z_{2n}Z_{2n+1}Z_{2n+2}
\bigr),
\end{align}
which satisfy
\begin{equation}
\prod_m X_m \, \mathcal{U}_{2n,2n+2}
=
-
\mathcal{U}_{2n,2n+2}
\, \prod_m X_m .
\end{equation}
Combining two such gates, we define a $\mathbb{Z}_2$-symmetric operator
$
\mathcal{W}_{2n,2n+4}
=
\mathcal{U}_{2n,2n+2}
\mathcal{U}_{2n+2,2n+4}.
$
The two states are therefore related by
$
\ket{\tilde{\Psi}_1}
=
\prod_{n\in\odd}
\mathcal{W}_{2n,2n+4}
\ket{\Phi_1},
$
provided that the system size is a multiple of $4$. 
Since each $\mathcal{W}$ gate is real, it also preserves $\mathbb{Z}_2^T$. 
Consequently, all gates of the entire circuit connecting  $\ket{\tilde{\Psi}_1}$ and $\ket{\Phi_1}$ respects the full $\mathbb{Z}_2 \times \mathbb{Z}_2^T$ symmetry. 
We therefore conclude that $\ket{\tilde{\Psi}_1}$ and $\ket{\Phi_1}$ are smoothly connected by a symmetry-preserving path and they are in the same phase. Consequently, ES of $\ket{\tilde{\Psi}_1}$ and $\ket{\Phi_1}$ share the same universal properties (except some fine-tuned points).   
Moreover, because ES of $\ket{\tilde{\Psi}_1}$ and $\ket{\Psi_1}$ are exactly the same, we conclude that ES of $\ket{\Psi_1}$ and $\ket{\Phi_1}$ share the same universal properties.

As a result, the gSPT state $\ket{\Psi_{\gSPT}(0)}$ in Eq.~\eqref{eq:psi_gSPT}, which contains a gapped sector, and the state $\ket{\Psi_{12}}$ in Sec.~\ref{Sec:Z_2T_and_c_half}, which does not, belong to the same $\mathbb{Z}_2 \times \mathbb{Z}_2^T$ gSPT phase.
By the same reasoning, one can show that the gSPT state $\ket{\Psi_{\gSPT}(\frac{\pi}{4})}$ in Eq.~\eqref{eq:psi_gSPT} and the state $\ket{\Psi_{23}}$ in Sec.~\ref{Sec:Z_2T_and_c_half} are also in the same $\mathbb{Z}_2 \times \mathbb{Z}_2^T$ gSPT phase.

\begin{figure*}
    \centering
    \includegraphics[width=\linewidth]{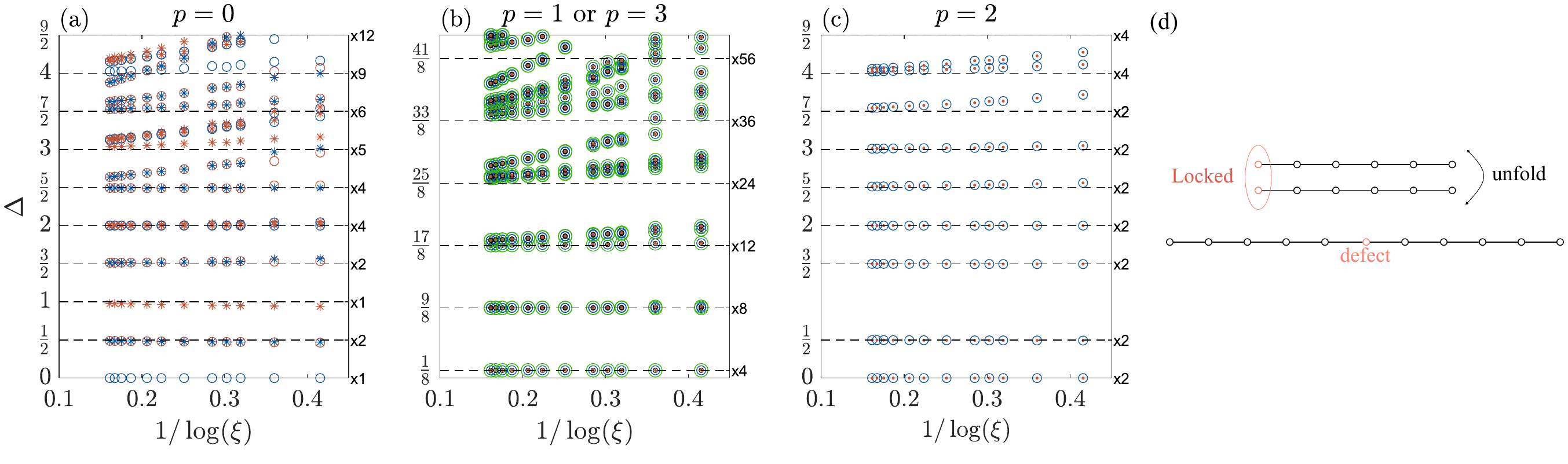}
    \caption{\textbf{Entanglement spectra (ES) of the $\mathbb{Z}_4 \times \mathbb{Z}_4$ gSPT states with $c=1$.}
All ES are extracted from the same set of $\mathbb{Z}_4$ symmetric iMPS (corresponding free physical boundaries) approximating the ground state of the critical $\mathbb{Z}_4$ clock model. The bond dimension $\chi$ ranges from $30$ to $600$.
(a) ES of the trivial gSPT state $\ket{\Psi_{0}}$, where different markers denote the $\mathbb{Z}_4$ charges of the iMPS.  
(b) ES of the non-trivial gSPT state $\ket{\Psi_{1}}$ (equivalently $\ket{\Psi_{3}}$), where different markers correspond to the eigenvalues of $\hat{Z}_N$, see Eq.~\eqref{eq:channel_Z4_1}.  
(c) ES of the non-trivial gSPT state $\ket{\Psi_{2}}$, where different markers correspond to the eigenvalues of $\hat{Z}_N^2$ , see Eq.~\eqref{eq:channel_Z4_2}.  
(d) Analysis of the effective EH of $\ket{\Psi_{2}}$, obtained by coupling two Ising chains at their left boundary and subsequently unfolding the system.}
    \label{Fig:Z4_ES_conserve}
\end{figure*}

\section{ES of $\mathbb{Z}_4\times\mathbb{Z}_4$ gSPT states with central charge $c=1$}\label{app:Z_4}

In this section, we discuss the ES of $\mathbb{Z}_4 \times \mathbb{Z}_4$ gSPT states described by a $c=1$ CFT compactified on an orbifold (i.e., identifying $\phi$ and $-\phi$) instead of a circle. Their ES can be naturally understood using the mapping between the $\mathbb{Z}_4$ clock model and two decoupled Ising chains.

The Hamiltonian of the trivial $\mathbb{Z}_q \times \mathbb{Z}_q$ gSPT state is~\cite{TSUI_2017}:
\begin{equation}\label{eq:clock}
    H_{0} = -\sum_{n=1}^{N-2} \hat{Z}^{\dagger}_{2n}\hat{Z}_{2n+2} - \sum_{n=1}^{2N-1} \hat{X}_n + \text{h.c.},
\end{equation}
where 
\begin{equation}
\hat{Z} = \sum_{p=0}^{q-1} e^{\frac{2\pi \mathrm{i} p}{q}} \ket{\hat{p}}\bra{\hat{p}}, \quad 
\hat{X} = \sum_{p=0}^{q-1} \ket{\hat{p}}\langle\hat{(p+1) \bmod q}|
\end{equation} 
are $\mathbb{Z}_q$ qudit operators. Hats are used to distinguish these operators and states from the $\mathbb{Z}_2$ Pauli operators and qubit states. Notice that compared to the $\mathbb{Z}_3$ clock model in Sec.~\ref{sec:non_inv_SPT}, here $\hat{X}$ and $\hat{Z}$ are swapped. The $\mathbb{Z}_q \times \mathbb{Z}_q$ symmetry is generated by 
$\prod_{n=1}^{N-2} \hat{X}_{2n}$ and $\prod_{n=1}^{N} \hat{X}_{2n-1}$.  

Similar to Eq.~\eqref{eq:Z2_Z2_c_half_gtri}, we adopt free physical OBC. On the odd sublattice, the Hamiltonian is trivially on-site, while on the even sublattice it reduces to the $\mathbb{Z}_q$ clock model at its critical self-dual point. Therefore, the ground state of $H_0$ can be written as 
\begin{equation}
\ket{\Psi_0} = \ket{\hat{+}}_\odd \otimes \ket{\Psi_{\clk}}_\even,
\end{equation}
where $\ket{\Psi_{\clk}}$ is the ground state of the clock model, and 
$\ket{\hat{+}}_\odd = \prod_{n=1}^{N} \ket{\hat{+}}_{2n-1}$ with 
$\ket{\hat{+}} = \frac{1}{\sqrt{q}}\sum_{p=0}^{q-1} \ket{\hat{p}}$ being the eigenstate of $\hat{X}$.  

It is known that there are $q$ distinct gapped $\mathbb{Z}_q \times \mathbb{Z}_q$ SPT states in 1D~\cite{SPT_cohomology_2013}, which can be obtained by applying an MPU to the trivial ones~\cite{Santos_2015}. Similar to the $\mathbb{Z}_2 \times \mathbb{Z}_2$ case, the MPU consists of generalized $CZ$ gates for $\mathbb{Z}_q$ qudits:  
\begin{align}
    \hat{U}_{CZ} &= \prod_{n} \hat{CZ}_{2n-1,2n-2} \prod_{n} \hat{CZ}^{\dagger}_{2n-1,2n}, \notag\\
    \hat{CZ}_{n,n+1}&\ket{\hat{p}_n,\hat{p}_{n+1}} = e^{\frac{2\pi i p_n p_{n+1}}{q}} \ket{\hat{p}_n,\hat{p}_{n+1}}.
\end{align}
The $q-1$ distinct non-trivial $\mathbb{Z}_q \times \mathbb{Z}_q$ gSPT states are then
\begin{equation}
\ket{\Psi_p} = \hat{U}_{CZ}^p \left(\ket{\hat{+}}_\odd \otimes \ket{\Psi_{\clk}}_\even \right), \quad p = 1,2,\dots,q-1.
\end{equation}

Let us focus on the case $q=4$. The gSPT states $\ket{\Psi_1}$ and $\ket{\Psi_3}$ are related by complex conjugation, so they have the same ES. Therefore, it suffices to consider the ES of $\ket{\Psi_0}$, $\ket{\Psi_1}$, and $\ket{\Psi_2}$. Given the reduced density matrix $\rho_0$ of $\ket{\Psi_0}$ at the entanglement cut between sites $N-1$ and $N$, the effective reduced density matrices $\rho_i$ of $\ket{\Psi_i}$ can be obtained by applying the corresponding quantum channels to $\rho_0$:
\begin{align}
\rho_1&\cong\rho_3\cong\varrho_1=\sum_{p=0}^3P^{[p]}_N\rho_0 P^{[p]}_N, \quad P^{[p]}=\ket{\hat{p}}\bra{\hat{p}},\label{eq:channel_Z4_1}\\
\rho_2&\cong \varrho_2=\frac{1+\hat{Z}_N^2}{2}\rho_0 \frac{1+\hat{Z}_N^2}{2}+\frac{1-\hat{Z}_N^2}{2}\rho_0 \frac{1-\hat{Z}_N^2}{2}.\label{eq:channel_Z4_2}
\end{align}

Let us first analyze the ES of $\ket{\Psi_0}$. The $\mathbb{Z}_4$ clock model can be mapped to two decoupled Ising chains. By introducing two sets of $\mathbb{Z}_2$ Pauli operators at each site, distinguished by the superscripts $0$ and $1$, one obtains the following on-site transformation~\cite{Exact_DQCP_2023}:
\begin{align}\label{eq:Z4_to_Z2_trans}
    \hat{Z}_n &= \frac{e^{-i\frac{\pi}{4}}}{\sqrt{2}}\left(Z^{[0]}_n + i Z^{[1]}_n\right), \notag\\
    \hat{X}_n &= X^{[0]}_n\left(\frac{1 + Z^{[0]}_n Z^{[1]}_n}{2}\right)
    + X^{[1]}_n\left(\frac{1 - Z^{[0]}_n Z^{[1]}_n}{2}\right),
\end{align}
which preserves the ES. Under this transformation, the trivial gSPT model is mapped to:
\begin{equation}
    H_0 = -\sum_{n=1}^{N-2}\!\left(Z^{[0]}_{2n} Z^{[0]}_{2n+2} + Z^{[1]}_{2n} Z^{[1]}_{2n+2}\right)
    - \sum_{n=1}^{2N-1}\!\left(X^{[0]}_n + X^{[1]}_n\right),
\end{equation}
where there are two decoupled Ising chains with free physical OBC on the even sublattice.
Since the entanglement boundary conditions of the two Ising chains are free, the effective EH of $\ket{\Psi_0}$ is given by the sum of two decoupled EH in Eq.~\eqref{eq:H_E_Ising}. The ES shown in Fig.~\ref{Fig:Z4_ES_conserve}a is fully consistent with this picture: it is simply the direct combination of two copies of the ES shown in Fig.~\ref{Fig:ES_Z2_Z2_c_half}a.

We now analyze the ES of the non-trivial gSPT states $\ket{\Psi_1}$ and $\ket{\Psi_3}$. From Eq.~\eqref{eq:Z4_to_Z2_trans}, the $\mathbb{Z}_4$ qudit states can be mapped to two $\mathbb{Z}_2$ qubit states as
\[
\ket{\hat{0}}=\ket{00}, \quad
\ket{\hat{1}}=\ket{10}, \quad
\ket{\hat{2}}=\ket{11}, \quad
\ket{\hat{3}}=\ket{01}.
\]
Together with the quantum channel in Eq.~\eqref{eq:channel_Z4_1}, this mapping implies that the effective EH of $\ket{\Psi_1}$ (or equivalently $\ket{\Psi_3}$) is given by the sum of two decoupled EH in Eq.~\eqref{eq:H_E_Ising_mix} with fixed entanglement boundary condition. The ES shown in Fig.~\ref{Fig:Z4_ES_conserve}b is fully consistent with this picture: it is simply the direct combination of two copies of the ES shown in Fig.~\ref{Fig:ES_Z2_Z2_c_half}b.

Let us now analyze the ES of $\ket{\Psi_2}$. Using the transformation in Eq.~\eqref{eq:Z4_to_Z2_trans}, the quantum channel in Eq.~\eqref{eq:channel_Z4_2} can be rewritten in terms of the $\mathbb{Z}_2$ Pauli operators as
\begin{equation}\label{eq:Z4_p2_channel_in_Z2_rep}
    \varrho_{2}
    = \frac{1 + Z^{[0]}_N Z^{[1]}_N}{2}\,\rho_0\,\frac{1 + Z^{[0]}_N Z^{[1]}_N}{2}
    + \frac{1 - Z^{[0]}_N Z^{[1]}_N}{2}\,\rho_0\,\frac{1 - Z^{[0]}_N Z^{[1]}_N}{2}.
\end{equation}
Since the effective EH of $\rho_0$ consists of two decoupled copies of the EH in Eq.~\eqref{eq:H_E_Ising}, the projectors in Eq.~\eqref{eq:Z4_p2_channel_in_Z2_rep} identify the left boundary spins of Ising chain $0$ and Ising chain $1$. Specifically, in the EH one has $Z^{[0]}_1 = \pm Z^{[1]}_1$ when the projector $\frac{1 \pm Z^{[0]}_n Z^{[1]}_n}{2}$ is applied. Consequently, the effective EH of $\ket{\Psi_2}$ describes two Ising chains coupled at the left boundary.
By unfolding the coupled chains—i.e., by relabeling $Z^{[1]}_n$ ($n>1$) as $Z_{-n+2}$ and $Z^{[0]}_n$ as $Z_n$ (and similarly for the $X$ operators; note that the effective EH lives on a different lattice with that of the reduced density matrices)—the effective EH of $\ket{\Psi_2}$ can be mapped to a single Ising chain with a defect at the center~\cite{OSHIKAWA_1997}:
\begin{equation}
    H_{E,2} \sim
    -\sum_{n=-L+2}^{L-1} Z_n Z_{n+1}
    -\sum_{\substack{n=-L+2 \\ n\neq 1}}^{L} X_n
    - h_1 X_1,
\end{equation}
as illustrated in Fig.~\ref{Fig:Z4_ES_conserve}d.
In Ref.~\cite{OSHIKAWA_1997}, the critical Ising chain with a defect was studied in detail. It was shown that the boundary entropy remains constant upon tuning the defect strength, corresponding to a line of RG fixed points with continuously varying boundary (or defect) criticality. Nevertheless, the numerical ES shown in Fig.~\ref{Fig:Z4_ES_conserve}c coincides with the OBC energy spectrum of a single Ising chain with free physical boundaries, whose spectrum shares the the same universal features with those of the ES in Fig. \ref{Fig:ES_Z2_Z2_c_half}c, suggesting that $h_1 = 1$. At present, however, we do not have a microscopic derivation of the exact value of $h_1$, and we leave this issue for future work.

%

\end{document}